\documentclass[reqno]{amsart}
\title{Towards a definition of quantum integrability}
\def \Hil {\mathcal{H}}

\newtheorem{theorem}{Theorem}

\newtheorem{proposition}{Proposition}
\newtheorem{definition}{Definition}
\newtheorem{lemma}{Lemma}

\usepackage{xy,graphicx,amssymb}
\xyoption{all}
\usepackage{ucs}
\usepackage[utf8x]{inputenc}

\def\goth{\mathfrak}

\def \Hil {\mathcal{H}}

\def \F {\mathcal{F}}

\def \Hil {\mathcal{H}}
\def \F {\mathcal{F}}

\def \R {\mathbb{R}}
\def \N {\mathbb{N}}

\def \C {\mathbb{C}}
\def\u {\mathfrak{u}}

\def \Tr {\mathrm{Tr}}

\newcommand{\pd}[1]{\frac{\partial }{\partial #1}}

\newcommand{\scalar}[1]{\langle #1 \rangle }

\author{Jes\'us Clemente-Gallardo}
\address{Instituto de Biocomputaci\'on y F\'\i sica de los Sistemas Complejos \\
  Universidad de Zaragoza \\ Corona de Arag\'on 42 \\ 50009 Zaragoza (SPAIN)\\
and \\
Departamento de F\'\i sica Te\'orica\\
Facultad de Ciencias \\
Universidad de Zaragoza \\
50009 Zaragoza (SPAIN) \\}
\email{jcg@unizar.es}
\author{Giuseppe Marmo}
\address{Dipartamento di F\'\i sica Teorica \\ Universit\'a Federico II and INFN
  Sezione di Napoli
  \\ Via Cintia \\ 80126 Napoli (ITALY)}
\email{marmo@na.infn.it}

\begin{document}

\maketitle

\begin{abstract}

We briefly review the most relevant aspects of complete integrability for
classical systems and identify those aspects which should be present in a
definition of quantum integrability.

We show that a naive extension of classical concepts to the quantum framework
would not work because all infinite dimensional Hilbert spaces are unitarily
isomorphic and, as a consequence, it would not be easy to define degrees of
freedom. We argue that a geometrical formulation of quantum mechanics might
provide a way out. 

\end{abstract}

PACS:03.65.Ca, 02.40.Yy, 02.30.Ik

\section{Introduction}

In classical Mechanics, completely integrable systems have been widely
studied in the past thirty years,  many aspects have been considered
and several approaches have been proposed (\cite{Fadeev2007,FadTak:1986,Vilasi,Magri:1978}). Many soliton
equations have been shown to admit a Hamiltonian formulation and to be
completely integrable.  Many completely integrable systems have been
shown to arise as reduction of ``simple'' systems, very often
associated with geodetical motions on Lie groups (\cite{MichRat:1998}). In this
respect, they have been shown to be associated with Lie-Scheffers
systems \cite{LieSche:1893,CarGrabMar:2000}, i.e. systems whose solutions admit
a superposition rule \cite{CarGrabMar:2000}. Thus we may safely say that
classical completely integrable 
systems are well understood. On the other hand, it is commonly
accepted that our description of the physical world should rely on Quantum
Mechanics. As Dirac puts it: ``\textit{classical mechanics must be a suitable
limit of quantum mechanics}''.

Inspired by this statement, it is quite natural to look for a
definition of quantum integrability whose appropriate limit would reproduce
the classical definition and the multifacets situations one knows for
classical systems, namely: multiHamiltonian descriptions, recursion
operators, Lax pairs and so on \cite{DeFilVilMar:84,Lax:1968,GelDorf:1982}. On
the mathematical side, the 
quantization of soliton models has given rise to the beautiful
structure of Yang-Baxter relations  and the related formulation of
quantum groups. 

However, there are several problems if one tries to implement in a
naive way the classical situation. Roughly speaking, the problems can
be reduced to the fact that in infinite dimension all Hilbert spaces
are isometrically isomorphic and therefore all $\C^*$--algebras of
bounded operators acting on them are isomorphic. In particular, the
idea of using the maximal set of commuting operators to imitate a
maximal set of commuting first integrals of the classical situation
does not work. The reason is a theorem by von Neumann  stating that
any two commuting operators can be written as functions of a third
one, i.e. out of the finite dimensional maximal set of commuting
operators we can find just one operator allowing us to express all the
others as functions of it (\cite{vonneumann_book}).

From this point of view, what is missing is a way to distinguish, say,
the Hilbert space of states of one particle from the Hilbert space of
the states of many particles. In other terms: in the quantum setting
it is not clear how to define the notion of degrees of
freedom. Kirillov \cite{Kir:1999}, for instance,
suggested the idea  of introducing the 
notion of ``functional dimension''.  In some sense, we need to
implement at the quantum level the notion of ``functionally
independent first integrals''.

Besides these problems, other problems arise from the need of a sound
definition of ``suitable classical limit'', i.e. to attach a
mathematically reasonable meaning to the physicist's limit $\hbar\to 0$
\cite{MarScoSimVen:2005}. 

At the moment, the way we consider this limit depends on the quantum
picture we deal with. For instance, the Schr\"odinger picture versus the
Heisenberg one, the first dealing with equations of motion on a
Hilbert space the second dealing with equations of motion on a
$\C^*$--algebra.  The two pictures are usually connected by the so
called $GNS$--construction \cite{dixmier}. At the classical level the first picture
corresponds to describe the dynamics as a symplectic vector field on a
symplectic manifold, while the second one would correspond to the
description of the dynamics as a Poisson derivation of a Poisson
algebra. 

By using this analogy we may consider the Hermitian structure on the
vector space of states as the mathematical structure corresponding to
the classical symplectic structure and similarly the Lie-Jordan
algebra structure on observables would replace the classical Poisson
structure. 

However, before entering more deeply into the analysis of these
analogies, let us consider what would be a reasonable approach to the
integrability of a quantum evolution equation. Given the operator $H$ in
some concrete realization, say the Schr\"odinger picture, we look for a
Lie algebra $\mathfrak{g}$ and a group $G$ that are ``naturally''
associated with $H$. Typically, this would mean that there is a
unitary representation of $G$ on the Hilbert space (the carrier space
of our system) such that $H$ can be identified with the action of an
element in the universal enveloping algebra $U(\mathfrak{g})$ of the Lie
algebra $\mathfrak{g}$. The 
structure of the unitary representations of $G$ and their
decomposition into irreducible ones then yields information about the
spectrum and the eigenfunctions of $H$. This approach relies on the
implicit assumption that the theory of unitary representations of
$G$ is sufficiently developed so that it may be used as a
non-commutative Fourier analysis (non-commutative Harmonic analysis on
$G$). Nowadays a lot is known for a large class of connected
finite-dimensional Lie groups (\cite{varadarajan2}).

Fundamental tools for this analysis are provided by Mackey's
imprimitivity theory, a generalization of techniques elaborated by
Weyl and von Neumann within quantum mechanics characterizing such
representations; and the Kirillov-Konstant correspondence between
representations and coadjoint orbits. For the use of Lie-algebraic
methods in the study of representations it is instrumental also the
use of the space of smooth vectors (this originated with coherent
states introduced by Schr\"odinger and further elaborated by Bargmann
and Nelson); which is a module for the associative algebra
$U(\mathfrak{g})$. 

The connection with biHamiltonian classical systems arises with the
introduction of biunitary operators. They are associated with
one-parameter groups of transformations  which are unitary 
for alternative Hermitian structures on the same topological vector
space of states (\cite{MarMorSimVen:2002,MarScoSimVen:2005}). The geometrical
formulation of Quantum Mechanics  
turns out to be very useful to identify the relevant mathematical
structures so that we may, more easily formulate what we mean by
alternative structures compatible with a quantum evolution.

Let us recall briefly the fundamental aspects of classical
integrability.

\section{Composition of model dynamics}

\subsection{Classical integrability}

From a classical point of view, the theory of integrable systems is very well
established.  In this section we shall just discuss a few of the best known
properties which will become later important when  compared to the analogue
situation in the quantum framework. 

Let us begin considering the usual definition of a classical integrable system.
Consider a classical system with $n$ degrees of freedom, defined on some
differential manifold $M$. For 
simplicity, let us consider a dynamics $\Gamma$ defined via a Hamiltonian
function $H$ on the cotangent bundle $T^*M$.  We will consider the
canonical Poisson structure associated to the canonical symplectic form, and
the corresponding Poisson bracket defined on $C^\infty(T^*M)$.

\begin{definition}
The system is said to be {\bf integrable} if there exists 
\begin{itemize}
\item a set of $n$
functions $\{ f_j\}_{j=1}^n$, where $f_k\in C^\infty(T^*M)$, which are constants of the motion
$$
0=\{ H, f_k\}  \qquad \forall k
$$
\item functionally independent $df_1\land \cdots \land df_n\neq 0$
\item which are in involution, i.e.
$$
\{ f_k, f_j\}=0 \qquad \forall j,k
$$
\end{itemize}
 
\end{definition}

Of course we could consider more general situations to describe systems
whose dynamics can be explicitly integrated. In principle, just the flow of
the system on the given manifold is to be considered as the object of interest.
The geometrical structures around it may be changed and actually
this leads to important consequences in some situations, as in the case of
biHamiltonian systems.  

There a few properties of these systems which we would like to consider more
closely. First of all, one of the most relevant properties is that the dynamics
for 
such a system  is always integrable by quadratures.  For instance, we can write 
the dynamics in a trivial way in a new set of coordinates, defined by the set
of the constants of the motion and their corresponding conjugate
variables. These can always be found, at least locally
\cite{arnold,abrahammarsden,MSSV:1985}:  

\begin{theorem}
There exists a set of coordinates $\{ I_j,\phi_j\}_{j=1}^n$ such that the
mapping $\Phi: (q^i, p_i)\mapsto (I_k, \phi_k)$ is a local symplectomorphism
and such that the vector field $\Phi_*(\Gamma)$ is written as

\begin{equation}
  \label{eq:nil}
\Phi_*(\Gamma)=\sum_k\omega_k(I)\pd{\phi_k}  
\end{equation}

\end{theorem}

This is one of the relevant features we would like to be able to recover at the
quantum level: the possibility of integrating completely the dynamics in a
constructive way. This is one of the more involved points in what regards the
definition of quantum integrability, because, within the quantum setting,  it
requires nonlinear 
transformations of non commuting variables, and we shall discuss it briefly. 

There is another aspect of classical integrability which we would like also to
mention here. The definition above is somehow restrictive from certain points
of view. If what we want is just to integrate the dynamics,  we could drop the
requirement that $\Phi$ be a symplectomorphism and just require that
(\ref{eq:nil}) is a vector field which is nilpotent of index two. This
would allow to cover much more general situations.  Let us review this quickly:  

\subsection{The meaning of Arnold-Liouville theorem}

        From our point of view, Arnold-Liouville's theorem (AL in the
        following) can also be read 
in a different way. One should interpret it as a way of relating a given 
system, defined in the original manifold $M$  with a set of systems defined on
suitable associated model manifolds $\{ N_j\} $ via ``projections'' $\Phi_j:M\to N_j$, defined
in action-angle variables, whose dynamics is trivially integrable. The role
of the $n$ constants of the motion in involution is just to give rise to 
the transformation which allow to define these ``projections''. This transformation is
nothing but the set of mappings defining each torus as a leaf of the
foliation of $M$ according to the level sets of the momentum mapping.
Each leaf may be therefore identified with  our manifold $N_j$, and the transformation
$$
\{(q_i,p^i)\}\mapsto (I_j, \phi_j)
$$
is the corresponding mapping $\Phi_j$. 

The aspect that we want to point out is that, from this perspective, AL
construction is analogous to the transformation of a linear system into Jordan
 blocks. The submanifolds characterized by  
$\Phi$ and $\nu$ are  now the analogues of the Jordan blocks. There is an
important difference however: AL theorem shows that by using nonlinear canonical
transformations we can relate generically any complete integrable systems with
two types of  models: 
\begin{itemize}
\item ``Harmonic oscillator''-type for those systems whose dynamics is completely
  bounded 
\item ``free particles''-type for those systems with unbounded motions.
\end{itemize}

        The decomposition of the dynamical vector field  is also obtained
as above.
Suppose that we are working with action-angle variables  on the
classical manifold $M$ that we assume to be symplectic. Classical dynamics is
described in terms of a vector field, which is the derivation associated
to the classical Hamiltonian $H$. Suppose we have an integrable system,
whose Hamiltonian vector field is written, in the usual way, in
action-angle variables $\{I_j,\phi_j\}_{j=1\cdots n}$ as:
$$
X_h=\nu_i\frac{\partial}{\partial \phi^i}
$$
where each $\nu_j\in \R$ is the corresponding ``frequency'' and we assume $\phi_j\in
(-\pi, \pi )$ (we assume hence each single dynamics will be bounded).  This vector
field describes the motion of a ``classical Harmonic 
oscillator'' and its integral curves are contained in the corresponding
toroidal manifold which serves as level set for the momentum mapping. On any
torus, the dynamical vector field has the same form, since the frequencies are
constant, however, in general, these constants may depend on the torus

{\bf Remark}

We should point out that by means of a nonlinear transformation we are going
from a system described, say,  by the matrix
$$
\left (
  \begin{array}{cc}
    0 & -1 \\
    1 & 0
  \end{array}
\right )
$$
in the $(q,p)$ coordinates to a system described by the nilpotent matrix
$$
\left (
  \begin{array}{cc}
    0 & 1 \\
    0 & 0
  \end{array}
\right )
$$
in the $(\Phi, \nu)$ coordinates. Thus we are considering a nonlinear transformation
which relates two linear dynamical systems described by completely different
matrices.

We may also describe the structure above in a generalized fashion by requiring
the existence of a set
of functionally independent functions $\{ J_1, \cdots , J_{2n}\} $ which satisfy 
$$
\mathcal{L}_\Gamma (\mathcal{L}_\Gamma J_j)=0,
$$
with $\Gamma $ being the dynamical vector field. We obtain in this way a description
in terms of generalized ``action-angle functions'', 
which need not be pairwise ``conjugate'', instead of the local ``action-angle
coordinates''. It 
means, in particular, that a completely integrable system is nilpotent of order
two, on some set of functionally independent functions which separate points of
the carrier space. In these ``coordinate functions'' the formal exponentiation
gives  
$$
e^{tL_\Gamma}J_j=J_j+tL_\Gamma J_j,
$$
i.e. the motion is linear in time. If we replace $J_j$ with $(I_j,\phi_j)$ we
recover the dynamics on each torus:
$$
I_j(t)=I_j(0) \qquad \phi_j(t)=\phi_j(0)+t\nu_j
$$

        Let us consider more closely integrable systems which are not
 described by isoperiodic motions.  The vector field on each torus is always of the same
type as above, but now the frequencies exhibit a dependence on the action
variables, i.e. $\nu_i=\nu_i(I_1,\cdots ,I_n)$. On each torus the
vector field is ``equivalent'' to the vector field of the Harmonic oscillator,
i.e. a linear combination of the angular vector fields, but the
coefficients are different for each torus. As a consequence,
we can say that a classical integrable system is made up by a suitable
combination of vector fields corresponding to Harmonic oscillators, where
suitable means that the oscillator
vector fields from each toroidal leaf are ``composed'' by using ``coupling
constant'' which are functions of the action variables.  

\subsection{Conjugate systems and classical integrability}
      In this section we want to elaborate further, with some additional
details and from a more geometrical approach, the  
interpretation of AL theorem that we just described.
The idea is that the integrability of a vector field $X$ (the dynamical
 vector field of our system) may be related with that of a set of models
 $\{Y_j\}$, related to a transversal decomposition of the original manifold,
whose evolution is well known. So we say that the integrability of each
one of the $Y_j$ ensures the integrability of the vector field $X$, which
is supposed to be decomposed in an analogous manner to the canonical ``Jordan form''
of a linear system. We will see in the following that this procedure offers
many advantages. 

        A crucial part of the present work concerns thus the relation between
structures belonging to different spaces, particularly vector fields and
algebras of functions. We will begin with the differential 
geometric preliminaries needed for the understanding of our considerations.

        Let us consider two differential manifolds, $M$ and $N$ and a 
differentiable mapping between them $\Phi:M\to N$
We suppose that we have a vector field on each manifold (the dynamical vector 
field for instance), and we want that $\Phi$ relates the different set of 
orbits on each space, in a 1:1 correspondence, without requiring that they have
the same parametrization. The general expression for
this correspondence is summarized  in the expression:

\begin{equation}
T \Phi \circ  X= f Y\circ  \Phi
\label{mapping_f}
\end{equation}

where $T\Phi$ is the induced tangent bundle  map acting on vector fields,
i.e. sections of $TM$ and $TN$ respectively,  and $f$ is an
arbitrary function on $N$. In other words, $X$ and $Y$ are $(\Phi,f)$--related,
i.e. $X$ and $fY$ are $\Phi$--related (this is a way to formalize the statement
that the orbits are in one-to-one correspondence up to parametrization). 

        When this mapping is such that $f=1$ we have a 1:1 
correspondence not only for orbits, but also for integral curves, i.e.
the image by $\Phi$ of any  integral curve  of $X$ on $M$ is an integral curve of
$Y$ on $N$; in this case we say simply that $X$ and $Y$ are $\Phi$--related.

        We will use now this kind of mappings, suitably combined, to decompose
the dynamics on a given manifold, defined through a Hamiltonian function or a
Hamiltonian vector field, in a set of well known dynamics on special manifolds
which we will call ``models''.

In fact we will not use the dynamics on the full manifold but just on an open
dense submanifold, extending suitably.

        Let us apply now some general considerations. Let
$M$ be a symplectic manifold (it is not necessary, but it will be so for
our applications) and let $X\in \goth{X}(M)$ be the dynamical vector
field. Let us call model systems $\mathcal{M}_j$, some pairs $(Y_j,N_j)$
where $N_j$ is a differential manifold and $Y_j\in \goth{X}(N_j)$. We
suppose we can define a  covering $\{ M_j\} $ of $M$ along with a set of
diffeomorphisms:
$$
\Phi_j:M_j\subset M\to N_j
$$
which posses property (\ref{mapping_f}).

        Now we want to use these sets $N_j$ and these vector fields
$Y_j$ to describe the dynamics corresponding to $X$. We can choose two
different ways for doing it: working at the level of vector fields and manifolds or
working at the level of functions and their algebraic structures:

\begin{itemize}
\item[a)] We can work at the level of vector fields, where we know the
dynamics is described by $X$, and we can try to find a suitable set of
vector fields $Y_j$ in such a way that we can integrate this dynamics
with their aid. We suppose that (\ref{mapping_f}) holds for any $N_j$, and that
a  decomposition may be chosen in such a way that we could write:
\begin{equation}
X=\sum_j h_j\widetilde Y_j
\label{decomposition}
\end{equation}
where $\mathcal{L}_{Y_i}h_j=0 \ \ \ \forall i\neq j$ and
$\widetilde Y_j\in \goth{X}(M)$ is
the vector field on $M$ $\Phi_j$--related with $Y_j$.

        In this set, we define:
\begin{definition}
We say that the model set $\mathcal{M}=\{(N_j,Y_j)\}$ gives a transversal
decomposition for $\goth{X}(M)$ when we can write:
$$
\goth{X}(M)(m)=\bigoplus_j \left [\goth{X}^\uparrow _j(N_j)\right ](m)
$$
which means that at any point of the manifold $M$ the  ``lifts'' of the
vectors fields from $N_j$ to $M$ define a basis of the corresponding tangent space
$T_mM$. 
\end{definition}

Thinking in terms of the $C^\infty (M)$--module structure of the set of vector
fields of $M$, it is also interesting to further analyze
this decomposition. We could say that the set of vector
fields $\{ \widetilde Y_j\}$, the lifted vector fields, defines a system of
generators for the module $\goth{X}(M)$ with coefficients in $C^\infty(M)$.

        When we work in classical dynamics, the physical observables are
 functions $C^\infty(M)$, and therefore we want to describe the previous
 construction in terms  of functions. We can restate the previous requirements
 in the form

$$
\mathcal{L}_{\widetilde Y_j}\Phi^*_j(f)=\Phi^*_j(L_{Y_j}f) \ \ \ \ \forall f\in
C^\infty(N_j)\,.
$$

%\placedrawing{diagram.lp}{Scheme of a transversal decomposition with $k$
%factors}{0.75}

\item[b)] We can also work at the level of functions on each manifold
with the algebra structure provided by the Poisson bracket. The dynamics
now is described with the Hamiltonian function and its adjoint action
inside the Poisson bracket:
$$\frac {df}{dt}=\{f,h\}$$

 The set of
models now is just specified by $\{N_j\}$ (and the corresponding mappings
$\Phi_j$), since we write the definition of transversal decomposition as:
\begin{definition}
Given $M$ and $X$ as above, and the set of models and mappings $(N_j,
\Phi_j)$ we say that they define a transversal decomposition of an open dense
submanifold of $M$ 
when the set $\bigoplus_j d\Phi^*_j(C^\infty(N_j))$ generates $\goth{X}^*(M)$ as a module (the set
of 1-forms on $M$). As above we are requiring that
$$
d\Phi_j^*(C^\infty(N_j))\cap d\Phi_i^*(C^\infty(N_i))=0\,,\quad i\neq j\,.
$$
\end{definition}

        And moreover, we say that a vector field
$X\in \goth{X}(M)$ is decomposable according to the set of models
$\{\Phi_j\}$ if
$$
\mathcal{L}_X \Phi^*_j(C^\infty(N_j))\subset \Phi^*_j(C^\infty(N_j))
$$
i.e. each subring $\Phi^*(C^\infty (N_j))$ of functions is stable under the action of
the dynamical vector field. 

It is important to remark  that both approaches are not trivially related, since the
set of Hamiltonian vector fields (which is the set arising in this latter one)
does not define a $C^\infty (M)$--module, and therefore there is no possible
transversal decomposition associated with them.
\end{itemize}

{\bf Remark}

        From this point of view, even non--Hamiltonian systems may be
considered integrable. In the first cases we have discussed objects
corresponding to the infinite dimensional Lie algebra of Hamiltonian vector
fields. But it is also possible to formulate things directly at the level of the
$C^\infty(M)$--module of vector fields as a whole.
Suppose that the dynamical field can be written as
a linear combination of  derivations that are inner when restricted to a torus, with
coefficients depending on the action variables. Using the preceding
formalism we say that in the combination above  we
do not need to impose that the functions $h_j$ belong to $\Phi^*_j(N_j)$
but just that:
\begin{equation}
\mathcal{L}_{\widetilde Y_j}h=0
\label{generalized}
\end{equation}
and therefore the derivations corresponding to $X$ need not be inner
(corresponding to a Hamiltonian on $M$) on the whole manifold, but that it
coincides on each torus with an inner derivation which may depend on the torus.
We could have a function $h_j=h_j(I_1,\cdots ,
I_n)$ which is constant on each torus, but that depends on  all of the
constants of the motion, the equation (\ref{generalized}) above implies
only that the Hamiltonian commutes with the function $h_j$ and therefore
its most general form is an arbitrary function of the constants of the
motion of the system.

To put it differently, we could use one-forms which are not closed to describe
vector fields by means of the symplectic structure. What we require is only
that the one-form is expressed in terms of action-angle variables. In different
terms, in the equations 
$$
\dot I^k=0 \qquad \dot \phi^k=\nu^k(I_1, \cdots , I_n),
$$ 
we may drop the requirement that $\nu^k$ are partial derivatives of a function
(the Hamiltonian function) \cite{AlekGrabmarMich:1997}.

%Consider also the notion of ``generalized constant of the motion'' as follows:
%assume there is a submanifold $N\subset M$ and a projection $\Phi:M\to N$ such that the
%dynamical vector field $\Gamma_M$ is $\Phi$--related to the null vector field on $N$. If
%$M$ is a symplectic manifold, 1then all functions $\Phi^*(C^\infty (N))\subset C^\infty (M)$ define
%Hamiltonian vector fields with the generalized constant of the motion $\Phi$. 

\subsection{Examples of classical integrable systems}

        Now we are able to give a few examples which illustrate  the
situation in the classical case. We can consider two families:
\begin{enumerate}
\item Systems whose orbits are on tori.
Model systems in this case would be just combinations of  one dimensional Harmonic
oscillator. Examples of this group would be:
\begin{itemize}
\item The $n$ dimensional Harmonic oscillator
\item The $q$-deformed oscillator, which would be the simplest non
trivial example.
\item The closed orbits of the Kepler problem.
\end{itemize}

\item Systems whose orbits are on cylinders (i.e. on $\R^k\times T^{n-k}$).
The model systems now have to combine the one dimensional Harmonic
oscillator and the free particle, to include the unbounded motion. As
examples of this case we could mention:
\begin{itemize}
\item The unbounded orbits of the Kepler problem.
\item The Calogero-Moser type potentials.
\end{itemize}
\end{enumerate}

\subsubsection{Example: the Harmonic oscillator}
        Let us develop now some of these examples. Let us begin with
systems defined on tori.

        For this kind of systems, the preceding description of Liouville
integrability is straightforwardly adapted. Let us take as a first
example the $n$ dimensional Harmonic oscillator:
$$H=\sum_k \frac 12 (p_k^2+\omega_kq_k^2) $$

        If we select as first integrals the partial energies, we can define:
$$I_j=h_j=\frac 12 (p_j^2+\omega_jq_j^2)$$
and the Hamiltonian is therefore written:
$$H=\sum_k h_k=\sum_kI_k$$

        Liouville-Arnold theorem provides us also with the angle variables,
and the conformal factor $F_k$ is obtained following the expression above
as:
\begin{equation}
f_k=\frac{\partial H}{\partial I_k}=\omega_k
\end{equation}

Obviously, the Harmonic oscillator turns out to be integrable, and its
dynamics linear.

        The following example is the first non trivial modification: the $q$-deformed
classical oscillators that can be found for instance in \cite{beppe}.
The Hamiltonian may be written as:
$$H=\omega \sum_k\frac{\sinh \hbar \alpha_k\alpha_k^*}{\sinh \hbar}$$
where $\alpha_k=\frac 1{\sqrt{2}}\left( \frac {ip_k}{\sqrt{\omega}}+
\sqrt{\omega}q_k \right )$ .

        The invariant now is the classical ``number operator'', but the
conformal factor will not be a numerical constant, but a constant of the
motion, i.e. a function of the
invariants:
$$I_k=\alpha_k\alpha^*_k$$
$$f_k=\frac{\partial H}{\partial I_k}=\omega \frac \hbar {\sinh \hbar}
\cosh \hbar \alpha_k \alpha_k^*$$

        The role played by this conformal factor is quite clear, it
represents a frequency that is not  constant on the phase space, but it
is constant on each manifold $I_k,\phi_k$.

\subsubsection{Conformal factors}
        On $\R^2$, with coordinates $(x,p)$ we consider the dynamical
system:
$$
\dot x=p f(x^2+p^2) \qquad \dot p=-xf(x^2+p^2)
$$

It is immediate that the orbits of this system are all circles with center
at the origin. We can integrate our system for any initial condition
$(x_0,p_0)$:
$$
\left (
\begin{array}{c}
x(t) \\
p(t)
\end{array}
\right )
=
\exp \left [ tf(x_0^2+p_0^2)
\left (
\begin{array}{cc}
0 &1 \\
-1 &0
\end{array}
\right )
\right ] 
\left (
\begin{array}{c}
x_0 \\
p_0
\end{array}
\right )
$$

        It is clear that we are using the Harmonic oscillator system
$$
\left \{ 
\begin{array}{cr}
\dot x=&\omega p \\
\dot p=&-\omega x
\end{array}
\right .
$$
to find our flow and replace the constant frequency $\omega$ with a
constant of the motion $f$. If, however, we look for a diffeomorphism
"conjugating" our starting system with the Harmonic oscillator, we would
find that no such a diffeomorphism can exist because we know that the
period of each orbit is invariant under diffeomorphisms, and our two
systems have different periods for each value of the radius. Therefore the
conformal factor will help in declaring our system ``conformally equivalent'' to
the Harmonic oscillator. Thus without the conformal factor, each  $f$ would
provide a different equivalence class.

\subsubsection{The Kepler problem}
        We consider now the case of the Kepler problem. In principle the idea
is simple, we have just to proceed as we did above. As a first point, the
action-angle variable have to be determined. Following \cite{beppe-vilasi}
we find that:
\begin{eqnarray}
J_1&=&-\sqrt{p_\theta^2+\frac{P_\phi^2}{\sin^2\theta}} +
mk\left ( \frac{2mk}r-\frac{p_\theta^2}{r^2}-\frac{p_\phi^2}{r^2\sin^2\theta}
-p_r^2 \right )^{-1} \\
J_2&=&\sqrt{p_\theta^2+\frac{P_\phi^2}{\sin^2\theta}}-p_\phi \\
J_3&=&p_\phi \\
\phi_1&=&-\frac {\sqrt{ -m^2k^2r^2+2mk(J_1+J_2+J_3)^2r-
(J_1+J_2+J_3)^2(J_2+J_3)^2}}{(J_1+J_2+J_3)^2} + \nonumber  \\
&& \arcsin \frac{mkr-(J_1+J_2+J_3)^2}
{(J_1+J_2+J_3)\sqrt{(J_1+j_2+J_3)^2-(J_2+J_3)^2}}\\
\phi_2&=&\phi_1-\arcsin \frac{[mkr-(J_2+J_3)^2](J_1+J_2+J_3)}
{\sqrt{(J_1+j_2+J_3)^2-(J_2+J_3)^2}} - \arcsin \frac{(J_2+J_3)\cos \theta}
{\sqrt{(J_2+J_3)^2-J_3^2}} \\
\phi_3&=&\phi_2+\arcsin \frac{J_3\cot \theta}{\sqrt{(J_2+J_3)^2-J_3^2}} +\phi.
\end{eqnarray}
Analogously, we can write the Hamiltonian
in terms of these variables:
\begin{equation}
H=-\frac {mk^2}{(J_1+J_2+J_3)^2}
\end{equation}

        From our previous discussion, we can compute now the conformal factor
for this case, simply as the derivative:
$$f_k=\frac{\partial H}{\partial J_k}$$
For this case there is just one conformal factor:
\begin{equation}
f_1=f_2=f_3=\frac {2mk^2}{(J_1+J_2+J_3)^3},
\end{equation}
and the related vector field acquires the form:
\begin{equation}
X=\frac {2mk^2}{(J_1+J_2+J_3)^3}\left ( \frac{\partial}{\partial \phi_1}+
\frac{\partial}{\partial \phi_2}+\frac{\partial}{\partial \phi_3}\right )
\end{equation}

Now we have to consider how to extend these structures to the quantum domain.
It is clear that the geometrical formulation of Quantum Mechanics allowing for
nonlinear transformations on the space of pure states, the space of rays or the
complex projective space, will be more appropriate to carry on our considerations.

\section{The problems of Quantum integrability}
\label{sec:probl-quant-integr}
Let us try to adapt to the quantum setting the results of classical
integrability we have analyzed. We recall that in order to be able to explicitly
integrate the system, the main properties and requirements
of AL theorem were
\begin{itemize}
\item the number of degrees of freedom of the system, which determines the
  dimension of the manifold where the system evolves, is equal to the number of
  functionally independent constants of the motion in involution.  
\item it is important to characterize the geometrical
  objects the system may be associated with. The reason for this is that the existence of
  different geometrical structures compatible with a given dynamics may
  provide new constants of the motion (as it happens with classical
  biHamiltonian   systems, for instance)
\item the notion of independence of the constants of the motion is crucial for
  them to be used as coordinates in the action-angle description (and
  therefore they must be functionally independent). 
\end{itemize}

All these properties, well known and controlled in the classical framework,
are not without ambiguities when transferred to the
quantum setting. Let us study them in some detail:

\subsection{The notion of degrees of freedom}

\label{sec:noti-degr-freed}

Let us now turn our attention towards the von Neumann theorem we mentioned in
the introduction. It reads as follows:
\begin{theorem}
  Given a Hilbert space $\mathcal{H}$ and two observables $A, B$, if they
  commute, there exists a third operator $C$ such that both $A$ and $B$ are
  functions of $C$.
\end{theorem}

Thus if we look for observables which commute among themselves and with the
Hamiltonian, as the naive generalization of AL theorem would suggest, we face
the problem of dealing with their independence since
the theorem we have quoted asserts that they will be functionally dependent on
just one operator $C$. We can not then 
relay on the number of degrees of freedom of the system, for this is a concept
which does not have a clear meaning in the usual quantum setting. Let us
elaborate a bit in this direction.

We are going to see how it is not possible to define a
reasonable notion of degree of freedom in the case of quantum
systems.  To this aim, we shall consider an example taken from a paper
by Weigert \cite{weigert:1992}  and further elaborated by Hietarinta \cite{Hiee:1989}.

Consider the case of a one-dimensional Harmonic oscillator
\begin{equation}
  \label{eqn:harm_osc}
  H=\hbar \omega\left (a^\dagger a+\frac 12\right )=\hbar \omega\left 
   (\hat n+\frac 12 \right).
\end{equation}
We recall that there is a countable set of eigenstates for the
number operator $\hat n$:
$$
\hat n|n\rangle=n |n\rangle \qquad n\in \N_0=\{ 0,1,2, \cdots \} 
$$
which constitute an orthonormal basis for the carrier Hilbert space of
states of our system. For the creation and annihilation operators we
have, in this basis, 
$$
a|n\rangle =\sqrt{n}|n-1\rangle \qquad a^\dagger
|n\rangle=\sqrt{n+1}|n+1\rangle, 
$$
along with 
$$
|n\rangle =\frac{(a^\dagger)^n}{\sqrt{n!}}|0\rangle.
$$

We would expect that this system, being one dimensional, should have only one constant of
the motion. However, we are going to show how it is possible to define
two constant of the motion in involution by using the following trick.

Let us replace the label $n$ by two indices $n_1$ and $n_2$, both
taking all non-negative integer values and related to $n$ by the
following relation
$$
n=n_1+\frac 12 (n_1+n_2)(n_1+n_2+1).
$$
We can list then a few correspondences:
\begin{eqnarray*}
n=0\longrightarrow (n_1,n_2)=(0,0) \quad n=1\longrightarrow (n_1,n_2)=(0,1)
\quad n=2\longrightarrow (n_1,n_2)=(1,0)\\ n=3\longrightarrow (n_1,n_2)=(0,2)
\quad n=4\longrightarrow (n_1,n_2)=(1,1)  
\end{eqnarray*}

Considering a two dimensional grid (in an analogous way we can repeat
the argument with three operators getting thus a cubic grid, etc) we
find the following diagram

\begin{center}
\includegraphics[width=9cm]{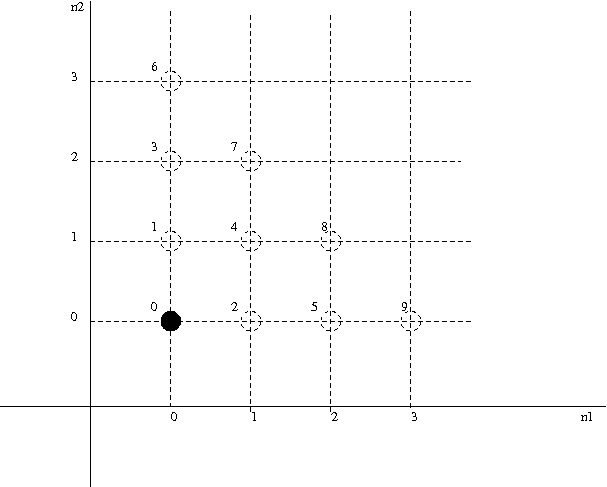}  
\end{center}

Every pair $(n_1,n_2)$ determines exactly one integer number $n$ and
vice-versa.  Therefore it is possible to label the states $|n\rangle$
by using the two integer-valued labels $(n_1, n_2)$:
$$
|n\rangle\equiv |n_1, n_2\rangle. 
$$
We can find, obviously, an orthonormal basis
$$
\langle n_1,n_2|n_1', n_2'\rangle=\delta_{n_1,n_1'}\delta_{n_2, n_2'},
$$
along with a decomposition of the identity
$$
\sum_{n_1, n_2}|n_1, n_2\rangle \langle n_1,n_2|=\sum_n|n\rangle
\langle n|=\mathbb{I}
$$
These states can be considered as the eigenstates of two operators,
$\hat n_1$ and $\hat n_2$, independent and hence commuting among
themselves. 

The eigenvalues of the Hamiltonian operator may be written as 
$$
E_n=\hbar \omega \left (n_1+\frac 12  (n_1+n_2)(n_1+n_2+1)
+\frac 12 \right) \qquad n_1, n_2\in \N_0
$$

The states $|0\rangle, |1\rangle , |3\rangle , \cdots |\frac
12k(k+1)\rangle$ belong to $n_1=0$ and therefore the eigenvalue has
infinite degeneracy. Similarly for $\hat n_2$, for which we can write:
$$
\hat n_2=\sum_{l=0}\sum_kl\hat P\left ( l+\frac 12 (k+l)(k+l+1\right).
$$  
Here $\hat P(m)=|m\rangle \langle m|$, i.e. this represents the operator giving
the spectral decomposition for $\hat n_2$. 

It is clear that $[\hat n_1, \hat n_2]=0$ and that the total
Hamiltonian operator can be written as
 
\begin{equation}
  \label{eqn:ham_n1n2}
\hat H=  \hbar \omega \left (\hat n_1+\frac 12  (\hat n_1+\hat
  n_2)(\hat n_1+\hat n_2+1)+\frac 12 \right) .
\end{equation}
Obviously, 
$$
[\hat H, \hat n_1]=0=[\hat H, \hat n_2].
$$

But this means that
\begin{lemma}
  The one dimensional Harmonic oscillator above has two constants of
  the motion commuting with each other.
\end{lemma}

We can consider the eigenstates $|n\rangle $ as belonging to the
Hilbert space which is the tensor product $\Hil_1\otimes \Hil_2$,
$\Hil_i$ being the Fock space corresponding to each number
operator. Both operators $\hat n_1$ and $\hat n_2$ have infinite
degeneracy, but the states of the composed operator $\hat n$ has none.  
$$
\hat H|k\rangle =k| k\rangle \quad k\in \N_0 \quad
|k\rangle=|n_1\rangle \otimes |n_2\rangle \quad n_1,n_2\in N_0.
$$

This has been done for the particular case of the Harmonic oscillator,
but as we are going to see, any one-dimensional quantum system (with
discrete spectrum and eigenvalues bounded from below) can be
given previous formulation. Consider for instance a system
described by the Hamiltonian
$$
\frac{\hat P ^2}{2m}+\hat V(q);
$$
and assume its spectrum is lower bounded as $E_0<E_1< E_2<\cdots
$. Denote the elements of the orthonormal basis of eigenstates as $\{
|E_i\rangle \}_{i=0, 1, \cdots }$. 

We can define now linear operators on the complex Hilbert space
spanned by the above vectors by setting:
$$
A|E_k\rangle =\sqrt{k}|E_{k-1}\rangle; \qquad A^\dagger |E_k\rangle =\sqrt{k+1}|E_{k+1}\rangle ,
$$
along with 
$$
A|E_0\rangle =0; \qquad \frac{(A^\dagger)^n}{\sqrt{n!}}|E_0\rangle .
$$

It is now obvious that the product $A^\dagger A\equiv \hat N$ has the
same eigenvectors as the original Hamiltonian $\hat H$ and thus $\hat
H$ can be written as a function of $\hat N$. Therefore the same
considerations we gave above for the Harmonic oscillator can be given
for a generic Hamiltonian (with the properties on the spectrum we
discussed above). But this implies that the system admits two
constants of the motion $\hat N_1$ and $\hat N_2$, in terms of which
the Hamiltonian can also be written.

More generally, given a maximal set of commuting operators, uniquely
identifying an orthonormal basis for the Hilbert space, once we order
these states by means of $\N_0$, we can define an operator
\begin{equation}
  \label{eqn:K}
  \hat K=\sum_k\frac{|k\rangle \langle k|}{(k+1)^2} \quad k\in \N_0.
\end{equation}
Thus, any element of the family of commuting operators will be an
appropriate function of $\hat K$.

It is clear that these peculiarities are to be traced in the fact that all
infinite dimensional Hilbert spaces are isomorphic, indeed unitarily isomorphic. If
$|n\rangle_{(a)}$ (with $n\in \N$) provides an orthonormal basis for
$\mathcal{H}_{(a)}$ (where $a=1,2,\cdots $), we can establish this correspondence by
setting $|n\rangle_2=T|n\rangle_1$. That being so, any operator $A_{(1)}$ on
$\mathcal{H}_{(1)}$ has a corresponding operator $A_{(2)}=VA_{(1)}V^{-1}$ on
$\mathcal{H}_{(2)}$. Thus also the corresponding $\C^*$--algebras are
isomorphic. This means that, at an abstract level, we can not implement a
notion of degree of freedom, because the Hilbert space of one particle is
isomorphic with the Hilbert space of many particles (Balachandran: bringing up
a quantum baby).

With these examples we hope to have convinced the reader that the
notion of degree of freedom in the framework of Quantum Mechanics does
not have, at the moment, a sound mathematical definition.

\subsection{The integration of the dynamics}

From the considerations we have made in Section \ref{sec:noti-degr-freed}, it
is clear that we need a 
notion of functional independence  for differentiable functions to be able to
distinguish the carrier space of one particle from the carrier space of many
particles. 

To the extent that the differential calculus is developed nowadays, to deal
with (local) derivations we need commutative algebras. If we start with
algebras of operators  on some Hilbert space, the spectral theorem  asserts
that every Hermitian operator is unitarily equivalent to a multiplication
function (operator). 

Thus, out of a family of commuting Hermitian operators we get a family of
multiplication functions. According to the Gelfand-Naimark theory, the point
spectrum of these commutative functions, with some additional requirements, can
be given the structure of a smooth manifold.  This seems to be the main route
to extract a ''classical manifold'' from a quantum carrier space.  (As
a matter of fact, the theory provides us with a Hausdorff topological space; to
get a smooth manifold one has to select a subset of states, often $C^\infty$ or
analytic vectors) \cite{ncg,landi_book,nelson}. 

In some other approaches these aspects may be hidden in other assumptions and
they appear in disguise. In many instances they appear very openly because the Hilbert
spaces  are required to be concretely realized  as square integrable functions
on some measure space, the algebra of operators is identified from the
beginning with differential operators (in this case the commutative algebra of
multiplications functions i.e. differential operators of degree zero,  is
immediately identified). A further choice of a 
specific quantum system, i.e. a specific Hamiltonian operator with a
corresponding domain of self-adjointness, will select further topological and
differential structures of the ''classical manifold''. 

In what follows we make few considerations on  a procedure which
subsumes most of the 
existing approaches to quantum integrability. We consider the equations of
motion on some Hilbert space $\mathcal{H}$, in the Schr\"odinger picture 
$$
i\hbar \frac{d}{dt}\psi=H\psi,
$$
or with the equations of motion on the algebra of operators (Heisenberg picture):
$$
i\hbar \frac{d}{dt}A=[H,A].
$$

We look for a Lie algebra $\mathfrak{g}$ and a related Lie group $G$ that are
''naturally'' associated with $H$. 

Typically, we look for a unitary representation of $G$ in the Hilbert space
$\mathcal{H}$ and require that the  Hamiltonian operator can be identified with
the action of an element in the universal enveloping algebra
$U(\mathfrak{g})$, $\mathfrak{g}$ being the Lie algebra of $G$. 

The structure of the unitary representations of $G$ and their decomposition in
irreducible ones yields information about the spectrum and the eigenfunctions
of $H$. Let us try to be more  specific by recalling a few facts about
representation theory: 

A unitary representation of $G$ in the Hilbert space $\mathcal{H}$ is a
homomorphism $\pi$ of $G$ into the group of unitary transformations
$U(\mathcal{H})$ with the following continuity condition:  

The map $G\times \mathcal{H}\to \mathcal{H}$ defined as  $(g,\psi)\mapsto \pi(g)\psi)$ is
continuous.  The homomorphism condition is equivalent to requiring that $\pi(g)$
is a continuous linear operator on $\mathcal{H}$ for all $g\in G$, that
$\pi(e)=\mathbb{I}$ ($e$ being the identity element of the group) and that the
product is preserved, i.e. $\pi(g_1g_2)=\pi(g_1)\pi(g_2)$ for all $g_1, g_2\in G$. 

If $\pi'$ is another representation  of $G$ in $\mathcal{H}'$, a continuous
linear map $T:\mathcal{H}\to \mathcal{H}'$ is called an {\bf intertwining
  operator} if $T\circ \pi(g)=\pi'(g)\circ T$ for all $g\in G$. This would be the ``quantum''
version of our ``related'' vector fields considered in the classical
setting. If $T$ is an isomorphism, $\pi$ 
and $\pi'$ are said to be isomorphic. Given a representation $\pi$, the continuous
operators $T:\mathcal{H}\to \mathcal{H}$ which intertwine $\pi$ with itself,
constitute an algebra, called the commuting algebra of the representation
$\pi$. For unitary representations, the isomorphic representations are unitarily
equivalent. 

For a unitary representation $\pi:G\to U(\mathcal{H})$, we may consider the set of
$C^\infty$ vectors (denoted as $\mathcal{H}^\infty$) and the set of analytic vectors
(denoted as $\mathcal{H}^\omega$). According to Nelson \cite{nelson}, both of them are 
always dense subspaces of the Hilbert space $\mathcal{H}$. Thus it is possible
to derive a representation of the Lie algebra $\mathfrak{g}$ of $G$ on these
subspaces by setting 
$$
\dot \pi:\mathfrak{g}\to \mathrm{End} (\mathcal{H}^\infty);
$$
given by
$$
\dot \pi(x) \psi=\lim_{t\to 0}\frac 1t \left (\pi(\exp (tx))\psi-\psi \right ).
$$

This representation has a natural extension to a representation of the
universal enveloping algebra $U(\mathfrak{g})$, which is still denoted by $\dot
\pi$.

If we define an involution on $U(\mathfrak{g})$ by setting 
$$
X^*=-X \qquad \forall X \in \mathfrak{g},
$$
the extension $\dot \pi$ to $U(\mathfrak{g})$ is a $*$--representation in the
sense that for all $Y\in U(\mathfrak{g})$ 
$$
\dot \pi (Y^*)\subseteq (\dot \pi(Y))^*.
$$

It is possible to introduce states on the universal enveloping algebra by
means of linear functionals. A functional $\Phi:U(\mathfrak{g})\to \C$ is said to be
a state if 
$$
\Phi(X^*X)\geq 0 \qquad \forall X\in U(\mathfrak{g}).
$$
This implies that $\Phi(X^*)=\bar \Phi(X)$ for all the elements $X\in
U(\mathfrak{g})$. Convex combinations of states are also states, and thus they
form a cone which defines a partial ordering on the linear space of all
self-adjoint linear functionals on $U(\mathfrak{g})$. Often, we consider
normalized functionals defined as $\Phi(\mathbb{I})=1$. 
The states that can not be written as convex combinations are extremal states
and are called pure states.

By using the Gelfand-Naimark-Segal (GNS) construction, with any state $\Phi$, we
associate a Hilbert space $\mathcal{H}_\Phi$, and a $*$--representation $\sigma$ of
$U(\mathfrak{g})$ by (unbounded) operators on a dense subspace $\mathcal{D}\subset
\mathcal{H}_\Phi$.  There exists also a vector $\Omega\in \mathcal{D}$ such that for all
$X\in U(\mathfrak{g})$,  
$$
\langle \sigma(x)\Omega|\Omega \rangle=\Phi(x),
$$
and $\mathcal{D}=\sigma(\{ U(\mathfrak{g})\}).\Omega$.

If $\Phi$ is an analytic state, the representation is an irreducible unitary
representation of the algebra if and only if $\Phi$ is a pure state. From the
construction, it is clear that by acting with the group $G$ on $\Omega$, we identify
a submanifold of states in $\mathcal{H}_\Phi$ which may be pulled back to $G$ and
therefore put into correspondence with a Hilbert space contained in
$\mathcal{L}^2(G)$ (the Hilbert space of square integrable functions on the
group with respect to the Haar measure). Thus, in conclusion, in this approach
a ``classical manifold'' appears again to be ``built in'' from the beginning
via the smooth manifold represented by the Lie group $G$. Let us mention that
the procedure considered in \cite{ErIbMarMor:2007} shows very clearly that by
using representations of $G$ on spaces  which carry alternative linear
structures would give rise to ``intertwining maps''  which are no more
linear. In this way the ``vector field'' point of view would not be ``optional''
any more but would be mandatory. Moreover this would be a way to implement
nonlinear transformations carrying from a Hilbert space description to another
Hilbert space description not linearly related among them. A situation like
this should be contemplated if we would like to have ``alternative Poisson
structures'' non necessarily compatible among them.

We shall mention now  few examples where the previous constructions can be
traced in the specific realizations  of groups and algebras.

\subsection{Finite level systems and alternative descriptions}

Within the geometrical formulation of Quantum Mechanics, a quantum
system is described by a vector field $\Gamma$ which preserves a
Hermitian structure. The carrier space, when considered not as a
complex vector space but as a real manifold, gives rise to a
Riemannian structure, a symplectic  structure and a connecting complex
structure (these aspects will be further elaborated in Section 4). A vector
field which describes quantum evolution has to 
preserve all of them: this means that it is not only symplectic, it is
also a Killing vector field and generates a complex linear
transformation. As the dynamics is linear and the structure above is
translationally invariant, we can represent the previous compatibility
conditions in terms of their representative matrices. 

We have
\begin{lemma}
  A matrix $A$ is the matrix associated with some linear Hamiltonian
  vector field if and only if it can be decomposed into the product of
  a skew symmetric matrix $\Lambda$ and a symmetric matrix $H$, i.e.
$$
A=\Lambda .H.
$$
\end{lemma}
Out of this decomposition we construct a Poisson bracket from $\Lambda$ and the Hamiltonian function from $H$.

If we consider now any linear invertible transformation $T:\Hil\to
\Hil$, we know the transformation law of the matrix $A$ representing a linear
transformation  ($A\mapsto
TAT^{-1}$) and we can write, by considering $\Lambda$ and $H$ as
representative of bilinear maps on account of the identification we  are going to make
\begin{equation}
  \label{eqn:lambda}
  A\mapsto A_T=TAT^{-1}=T\Lambda T^{t}(T^t)^{-1}HT^{-1}=\Lambda_TH_T.
\end{equation}
We have considered $\Lambda$ to transform contravariantly and $H$ to
transform covariantly.

If $T$ is a symmetry transformation for $A$, i.e $A_T=A$, we conclude
$$
A=\Lambda_TH_T,
$$
where, in general, $\Lambda_T$ will be different from $\Lambda$.  If
$\Lambda$ represents a Poisson structure, we would say that the
transformation $T$ is a symmetry for the dynamics which is not
``canonical''.  Thus the symmetry group for $A$, quotiented by the
subgroup of canonical transformations for $\Lambda$ would parametrize
the family of alternative decompositions for $A$ and consequently of
the family of alternative Hamiltonian descriptions
\cite{MorFerLoVecMarRu:1990}. 

There are a few properties of these alternative descriptions which are
immediate to list:
\begin{itemize}
\item all odd powers of $A$ are traceless, i.e.
$$
\Tr A^{2k+1}=0,
$$
and give rise to new Hamiltonian systems.
\item when $A$ is generic, the commutant and the bicommutant coincide
  and are generated by all powers of $A$.
\item in the generic case
$$
e^{\lambda A^2}\Lambda e^{-\lambda(A^T)^2}=\Lambda_\lambda, 
$$
for different values of  $\lambda$ gives representative matrices
of alternative Poisson structures.
\item in the quantum case, if $T$ in addition commutes with the
  complex structure (i.e. it is the real form of a complex linear
  transformation), then it will also generate alternative Hermitian
  structures when $H$ is positive definite.  In particular, in the generic
  case, $e^{\lambda A^2}$ 
  will generate alternative Hermitian structures (if we write $A= iB$
  where $B$ is Hermitian, it is evident that the even powers of $A$
  are not skew-Hermitian).
\end{itemize}

By using collective coordinates, say $\{ \xi_j\} $, for the
realification of the complex vector space, we have for the
corresponding tensor fields:
$$
\Gamma=\xi_kA^k_j\pd{\xi_j}; \qquad
\Lambda_\lambda=(\Lambda_\lambda)_{jk}\pd{\xi_j}\land \pd{\xi_k };
\qquad
H_\lambda=\frac 12 H_\lambda^{jk}\xi_j\xi_k,
$$
which explain the transformation properties we have used for $\Lambda$ and $H$.
We obtain thus alternative Hamiltonian descriptions.  It is important
to remark that the alternative Poisson structures are mutually
compatible, i.e. any combination of them with real valued coefficients
is still a Poisson structure.

To be more definite, we shall consider a specific example given by a
quantum system with a 2--dimensional Hilbert space  (i.e. a two-level
quantum system). The equations of the motion for such a system would
be:
$$
\dot S=[H,S] \qquad S=S_1\sigma_1+S_2\sigma_2+S_3\sigma_3 \qquad
H=B\sigma_3 
$$

The matrix $\sigma_3$ may be considered to be the representative
matrix of our dynamics, thus
$$
A=i
\left (
\begin{array}{cc}
 1 & 0 \\
0 & -1 
\end{array}
\right )
\quad
A^2=
\left (
\begin{array}{cc}
 -1 & 0 \\
0 & -1 
\end{array}
\right )
\quad
A^4=
\left (
\begin{array}{cc}
 1 & 0 \\
0 & 1 
\end{array}
\right )
$$

This case is not generic, and therefore we have to consider matrices in the
commutant of $\sigma_3$ which are skew Hermitian. Without going through the full
procedure, it is not difficult to see that all alternative Hamiltonian
descriptions may be written in the form
$$
h=a_1^2z_1\bar z_1+a_2^2z_2\bar z_2 \qquad a_1, a_2\in \R;
$$
and in real coordinates ($z_j=q_j+ip_j$)  we would have
$$
h=a_1^2(p_1^2+q_1^2)+a_2^2(p_2^2+q_2^2) \qquad a_1, a_2\in \R;
$$
along with corresponding symplectic and Riemannian structures
$$
g=a_1^2(dp_1^2+dq_1^2)+a_2^2(dp_2^2+dq_2^2) \qquad
\omega=a_1^2dp_1\land dq_1+a_2^2dp_2\land dq_2
$$

A further generalization arise if we let $a_1$ to become a function of
$z_1\bar z_1$ and $a_2$ a function of $z_2\bar z_2$ (see
\cite{DubMarSi:1990}), or more generally
$$
h=a_1^2(z_1\bar z_1, z_2\bar z_2)z_1\bar z_1+z_2\bar z_2a_2^2(z_1\bar
z_1, z_2\bar z_2) .
$$
In this case the 
skew-symmetric part of the Hermitian tensor is still given by a
symplectic structure, but the Riemannian part is no longer flat. If we
use a Darboux chart for the new symplectic structure, we can define a new
flat Riemannian tensor, and hence a new Hermitian structure, but 
it will not be connected to the original one by nonlinear ``point
transformations'', i.e. transformations only on the ``coordinates''. 
To put it differently, the Darboux coordinates are adapted to the symplectic
structure but are ``not adapted'' to the Riemannian tensor.
 The new
Darboux chart allows to introduce a new 
linear structure such that the original dynamics is still linear with
respect to the new linear structure. On each symplectic plane the
metric tensor is going to be
$$
(d\rho_1)^2 +\rho_1^2(d\phi_1)^2
$$
in the flat case, and there will be a conformal factor instead when
$a_j$ is a function of the coordinates.

By using the new linear structure associated with the Darboux chart,
it is possible to define a new realization of the unitary group $U(2)$
which is not linearly related to the original one \cite{ErIbMarMor:2007}.

To be able to make sense of some of these nonlinear manipulations, it is
convenient to consider a geometrical formulation of Quantum Mechanics.

\section{Geometric Quantum Mechanics}
The aim of this section is to present a brief summary of the set of the
geometrical tools which characterize the description of Quantum Mechanics. For
further details we refer the reader to
\cite{Grabowski:2005,Grabowski:2006,CaC-GMar:2007,CaC-GMar:2007b,C-GMar:2007,C-GMar:2008}.  

\subsection{The states}

To introduce the real manifold point of view, we start by replacing the Hilbert
space $\mathcal{H}$ with its realification $\mathcal{H}_\R$.
In this realification process the complex structure on
$\Hil$ will be represented by a tensor $J$ on $\Hil_\R$ as we will see.

The natural identification is then  provided by
$$
\psi_R+i\psi_I=\psi \in \Hil \mapsto (\psi_R,\psi_I)\in \Hil_\R.
$$
Under this transformation, the Hermitian product becomes, for $\psi^1, \psi^2\in
\Hil$ 
$$
\langle(\psi^1_R,\psi^1_I), (\psi^2_R,\psi^2_I)\rangle =(\langle \psi^1_R, \psi^2_R\rangle + \langle \psi^1_I,
\psi^2_I\rangle )+i( \langle \psi^1_R, \psi^2_I\rangle -\langle \psi^1_I,\psi^2_R\rangle ).
$$

To consider $\Hil_\R$ just as a real differential manifold, the algebraic
structures available on $\Hil$ must be converted into tensor fields on
$\Hil_\R$. To this end we introduce the tangent bundle $T\Hil_\R$ and
its dual the cotangent bundle $T^*\Hil_\R$. The linear structure available in
$\Hil_\R$ is encoded in the vector field $\Delta$
$$
\Delta:\Hil_\R\to T\Hil_\R \quad \psi \mapsto (\psi, \psi)
$$

We can consider the Hermitian structure on $\Hil_R$ as an Hermitian
tensor on $T\Hil_\R$. With every vector we can associate a vector field
$$
X_\psi:\Hil_\R\to T\Hil_\R \quad \phi \to (\phi, \psi)
$$
These vector fields are the infinitesimal generators of the vector
group $\Hil_\R$ acting on itself. 

Therefore, the Hermitian tensor, denoted in the same way as the scalar
product will be
$$
\langle X_{\psi_1}, X_{\psi_2}\rangle(\phi) =\langle \psi_1, \psi_2\rangle, 
$$
i.e. the tensor field we are defining is translationally invariant.

 On the ``real manifold'' the Hermitian scalar product may be  written as 
$
\langle \psi_1, \psi_2\rangle=g(X_{\psi_1}, X_{\psi_2})+i\,\omega (X_{\psi_1}, X_{\psi_2}),
$
where $g$ is now a symmetric tensor and $\omega $ a skew-symmetric one.  It is also
possible to write the associated quadratic form as a pull-back by
means of the dilation vector field $\Delta$ extended as  a map in an obvious way
$\mathcal{H}_R\to 
T\mathcal{H}_\R\oplus T\mathcal{H}_\R$: 
$$
(\Delta^*(g+i\omega))(\psi, \psi)=\langle \psi, \psi \rangle_ {\Hil} 
$$

The properties of the Hermitian product ensure that:
\begin{itemize}
\item the symmetric tensor is positive definite and non-degenerate, and hence
  defines a Riemannian structure on the real vector manifold.
\item the skew-symmetric tensor is also non degenerate, and is closed with
  respect to the natural differential structure of the vector space. Hence, the
  tensor is a symplectic form.
\end{itemize}

As the inner product is sesquilinear, it satisfies
$$
\langle \psi_1, i\psi_2\rangle =i\langle \psi_1, \psi_2\rangle, \qquad \langle i\psi_{1}, \psi_{2}\rangle =-i\langle
\psi_{1},\psi_{2}\rangle. 
$$
This  implies
$$
g(X_{\psi_1}, X_{\psi_2})=\omega (JX_{\psi_1}, X_{\psi_2}).
$$
We also have that $J^2=-\mathbb{I}$, and hence that the triple $(J, g, \omega )$
defines a K\"ahler structure.  This implies, among other things, that the tensor
$J$ generates both finite and infinitesimal transformations which are
orthogonal and symplectic.

Linear transformations on the vector space $\Hil_\R$ are converted
into $(1,1)$--tensor fields by setting $A\to
T_A$ where 
$$
T_A:T\Hil_\R\to T\Hil_\R \quad (\psi, \phi)\mapsto (\psi, A\phi).
$$
The association $A\to T_A$ is an associative algebra isomorphism. 
The Lie algebra of vector fields obtains by setting $X_A=T_A(\Delta)$.
Complex linear transformations will be represented by $(1,1)$--tensor fields
commuting with $J$, or by vector fields $Y$ such that $L_YJ=0$.

For finite dimensional Hilbert spaces it may be convenient to
introduce adapted coordinates on $\Hil$ and $\Hil_\R$.
An orthonormal basis $\{ |e_k\rangle \} $ of the Hilbert space allows us to
identify this product with the canonical Hermitian product on $\C^n$:
$$
\langle \psi_1, \psi_2\rangle =\sum_k\langle \psi_1, e_k\rangle \langle e_k, \psi_2\rangle 
$$
The group of unitary transformations on $\Hil$ becomes identified with the
group $U(n, \C)$, its Lie algebra $\mathfrak{u}(\Hil)$ with $\mathfrak{u}(n,
\C)$ and so on.

The choice of the basis also allows us to introduce coordinates for the
realified structure:
$$
\langle e_k, \psi\rangle =(q_k+ip_k)(\psi),
$$ 
and write the geometrical structures introduced above as:
$$
J=\partial_{p_k}\otimes dq_k-\partial_{q_k}\otimes dp_k 
\quad
g=dq_k\otimes dq_k+dp_k\otimes dp_k
\quad
\omega=dq_k\land dp_k
$$

If we combine them into complex coordinates we can write the Hermitian structure
by means of $z_n=q_n+ip_n$:
$$
h=\sum_kd\bar z_k\otimes dz_k
$$

In an analogous way we can consider a contravariant version of these
tensors. It is also possible to build it by using the isomorphism
 $T\Hil_\R\leftrightarrow
T^*\Hil_\R$ associated to the Riemannian tensor $g$. The result in
both cases  is a K\"ahler structure for the dual vector space $\Hil_\R^*$ with the dual
complex structure $J^*$, a Riemannian tensor $G$ and a (symplectic) Poisson
tensor $\Omega$: The coordinate expressions with respect to the natural basis are:
\begin{itemize}
\item the Riemannian structure  $
G=\sum_{k=1}^n\left(\pd{q_k}\otimes \pd{q_k}+\pd{p_k}\otimes \pd{p_k}\right),
$ 
\item  the Poisson tensor
$
\Omega=\sum_{k=1}^n\left(\pd{q_k}\land \pd{p_k}\right)
$
\item while the complex structure has the form 
$$J=\sum_{k=1}^n\left(\pd{p_k}\otimes d{q_k}-\pd{q_k}\otimes d{p_k}\right)$$
\end{itemize}

\subsection{The observables}
The space of observables (i.e. of self-adjoint operators acting on $\Hil$) may
be identified with the dual $\mathfrak{u}^*(\Hil)$ of the real Lie algebra
$\mathfrak{u}(\Hil)$, according to the pairing between the unitary Lie algebra
and its dual given by
$$
A(T)=\frac i2 \Tr AT
$$

Under the previous isomorphism, $\mathfrak{u}^*(\Hil)$ becomes a Lie algebra
with product  defined by
$$
[A, B]=\frac \hbar{2i}(AB-BA)
$$

We can also consider the Jordan product:
$$
[A,B]_+=2(A\circ B)=AB+BA
$$
with the associative product given by
$$
AB=A\circ B+\frac i\hbar [A, B],
$$
and defined on the complexified space $\mathfrak{gl}(\mathcal H)$.

Both structures are compatible in the sense that the Jordan product is
invariant under the linear transformations on $\u^*(\Hil)$ defined by
the adjoint action (of the Poisson product). As a result,
$\mathfrak{u}^*(\Hil)$ becomes a 
Lie-Jordan algebra (see \cite{Emch,Lands:book}).

We can also define a suitable scalar product on Hermitian operators, given by:
$$
\langle A, B\rangle=\frac 12 \Tr AB
$$
which turns the real vector space into a real Hilbert space. This scalar product is the
restriction of the one on $\mathfrak{gl}(\Hil)$ defined as $\langle M, N\rangle =\frac 12
\Tr M^\dagger N$.

Moreover this scalar product is compatible with the Lie-Jordan structure in the
following sense:
$$
\langle [A,\xi], B\rangle _{\mathfrak{u}^*(\Hil)}=\langle A, [\xi, B]\rangle _{\mathfrak{u}^*(\Hil)}
\quad
\langle [A,\xi]_+, B\rangle _{\mathfrak{u}^*(\Hil)}=\langle A, [\xi, B]_+\rangle _{\mathfrak{u}^*(\Hil)}
$$

These algebraic structures may be given a tensorial formulation in terms of the
association $A\mapsto T_A$ on the tangent bundle
$T\u^*(\Hil)$. However we can also associate complex valued functions 
with linear operators $A\in \mathfrak{gl}(\Hil)$ by means of the scalar product
$$
\mathfrak{gl}(\Hil)\ni A\mapsto f_A=\frac 12 \langle \psi, A\psi\rangle_{\Hil}. 
$$

In more intrinsic terms we may write
$$
f_A=\frac 12 (g(\Delta, X_A)+i\omega(\Delta, X_A)).
$$
Hermitian operators give rise thus to quadratic real valued functions.

The association of operators with quadratic functions allows also to recover
the product structures on $\u(\Hil)$ and $\u^*(\Hil)$ by means of
appropriate $(0,2)$--tensors on $\Hil_\R$. Indeed, by using the contravariant form of
the Hermitian tensor $G+i\Omega$ given by: 
$$
G+i\Omega=4 \frac{\partial}{\partial{z_k}}\otimes \frac{\partial}{\partial{\bar
  z_k}}=\frac{\partial}{\partial{q_k}}\otimes \frac{\partial}{\partial{q_k}}+
\frac{\partial}{\partial{p_k}}\otimes \frac{\partial}{\partial{p_k}}+
i \frac{\partial}{\partial{q_k}}\land \frac{\partial}{\partial{p_k}},
$$

it is possible to define a bracket
$$
\{ f,h\} _{\Hil}=\{ f,h\}_{g}+i\{ f,h\} _{\omega}
$$

We remark that $\frac{\partial}{\partial{z_k}}$ and $\frac{\partial}{\partial{\bar
  z_k}}$ are not to be considered as derivations with respect to the complex
coordinates introduced above but as complex valued smooth vector fields.

In particular, for quadratic real valued functions we have
$$
\{ f_A,f_B\}_g =f_ {AB+BA}=2f_{A\circ B} \quad
\{ f_A,f_B\}_{\omega}=-if_{AB-BA}
$$

The imaginary part, i.e. $\{ \cdot, \cdot\} _\omega$, defines a Poisson bracket on the space of
functions. Both brackets  define a tensorial version of the
Lie-Jordan  algebra for the set of operators.

For Hermitian operators we recover previously defined vector fields:
$$
\mathrm{grad}f_A=\widetilde A; \quad \mathrm{Ham} f_A=\widetilde {iA}
$$
where the vector fields associated with operators, we recall,  are defined by:
$$
\widetilde A:\Hil_{\R}\to T\Hil_{R} \quad \psi\mapsto (\psi, A\psi)
$$

$$
\widetilde{iA}:\Hil_\R\to T\Hil_{R} \quad \psi\mapsto (\psi, JA\psi) 
$$

We can also consider the algebraic structure associated to the full
 bracket $\{
\cdot, \cdot \}_\Hil $, as we associated above the Jordan product and 
the commutator of operators to the brackets $\{ \cdot, \cdot\} _g$ and $\{ \cdot, \cdot\} _\omega$
respectively. It is 
simple to see that it corresponds to the associative product of the set of
operators, i.e.
$$
\{ f_A, f_B\} _\Hil=\{ f_A, f_B\} _g+i \{ f_A, f_B\} _\omega=f_{AB+BA}+f_{AB-BA}=
2f_{AB}
$$

This particular bilinear product on quadratic functions may be used to define
a star product on quadratic functions
$$
\{ f_A, f_B\} _\Hil=2f_{AB}=\langle df_A, df_B\rangle_{\Hil^*}=f_A\star f_B
$$ 
The set of quadratic functions endowed with such a product turns out to be a
$\C^*$--algebra.

Summarizing, we can reconstruct all the information of the algebra of
operators starting only with real-valued functions defined on $\Hil_\R$. We
have thus
\begin{proposition}[\cite{CirManPizzo:1994}]
  The Hamiltonian vector field $X_f$ (defined as $X_f=\hat \Omega(df)$) is a Killing
  vector field for the Riemannian tensor $G$ if and only if $f$ is a quadratic
  function associated with an Hermitian operator $A$, i.e. there exists $A=A^\dagger
  $ such that $f=f_A$. 
\end{proposition}

Finally, we can consider the problem of how to recover the eigenvalues and
eigenvectors of the operators at the level of the functions of
$\Hil_R$. We consider the expectation value functions associated to
the operators as: 
$$
 A\mapsto e_A(\psi)=\frac{\langle \psi, A\psi\rangle}{\langle \psi, \psi \rangle}. 
$$

Then,
\begin{itemize}
\item eigenvectors correspond to the critical points of functions $e_A$, i.e.
$$
de_A(\psi_a)=0 
$$
iff  $\psi_a$ is an eigenvector of $A$.
We notice that the invariance of $e_A$ under multiplication by a phase
$U(1)$ implies that critical points form a circle on the sphere of normalized
vectors if the eigenvalue is not degenerate.  
\item the corresponding eigenvalue is recovered by the value 
$
e_A(\psi_a)
$
\end{itemize}

Thus we can conclude that the K\"ahler manifold $(\Hil_\R, J, \omega, g)$
contains all the information of the usual formulation of Quantum Mechanics on a
complex Hilbert space.

Up to now we have concentrated our attention on states and observables. If we
consider observables as generators of transformations, i.e. we consider the
Hamiltonian flows associated to the corresponding functions, the invariance of
the tensor $G$ implies that the evolution is actually unitary. It is,
therefore, natural, to consider the action of the unitary group on the
realification of the complex vector space.

\subsection{The momentum map: geometrical structures on 
$\mathfrak{u}^*$} 
The unitary action of $U(\Hil)$ on $\Hil$ induces a symplectic action on the
symplectic manifold $(\Hil_\R, \omega)$.  By using the association
$$
F:\Hil_\R\times \u(\Hil)\to \R \quad (\psi, A)\mapsto \frac 12 \langle \psi, iA\psi\rangle =f_{iA}(\psi),
$$
we find, with $F_A=f_{iA}:\Hil_\R\to \R$, that
$$
\{ F(A), F(B)\}_\omega =iF([A,B]).
$$
Thus if we fix $\psi$, we have a mapping $F(\psi):\u(\Hil)\to \R$. With any
element $\psi \in \Hil$ we associate an element in $\u^*(\Hil)$. The previous
map defines a momentum map  
$$
\mu:\Hil\to \mathfrak{u}^*(\Hil),
$$
which provides us with a symplectic realization of the natural Poisson manifold
structure available in $\u^*(\Hil)$. 
We can write the momentum map from $\Hil_\R$ to $\mathfrak{u}^*(\Hil)$ as
$$
\mu(\psi)=|\psi\rangle \langle \psi|
$$

If we make the convention that the dual
$\u^*(\Hil)$ of the (real) Lie algebra $\u(\Hil)$ is identified with Hermitian
operators by means of a scalar product, the product pairing between Hermitian
operators $A\in \u^*(\Hil)$ and the anti-Hermitian element $T\in \u(\Hil)$ will be
given by
$$
A(T)=\frac i2 \Tr (AT)
$$

If we denote the linear function on $\u^*(\Hil)$ associated with the element
$iA\in \u(\Hil)$ by $\hat A$, we have
$$
\mu^*(\hat A)=f_A
$$

The pullback of linear functions on $u^*(\Hil)$ is given by the quadratic
functions on $\Hil_\R$ associated with the corresponding Hermitian operators.

It is possible to show that the contravariant tensor fields on $\Hil_\R$
associated with the Hermitian structure are $\mu$--related with a complex
tensor on $\u^*(\Hil)$:
$$
\mu_*(G+i\Omega)=R+i\Lambda; 
$$
where the two new tensors $R$ and $\Lambda$ are defined by
$$
[R(\hat A, \hat B)](\xi)=\scalar{\xi, [A,B]_+}_{\u^*}=\frac 12 \Tr (\xi (AB+BA))
$$
and
$$
[\Lambda(\hat A, \hat B)](\xi)=\scalar{\xi, [A,B]_-}_{\u^*}=\frac 1{2i} \Tr (\xi (AB-BA))
$$

Clearly,
$$
G(\mu^*\hat A, \mu^*\hat B)+i\Omega(\mu^*\hat A, \mu^*\hat B)=\mu^*(R(\hat A, \hat
B)+i\Lambda(\hat A, \hat B)).
$$

%As we know that $\u^*(\Hil)$  is foliated by symplectic manifolds, we wish to
%consider more closely the map from  $\Hil_\R$ to the minimal symplectic orbit
%on $\u^*(\Hil)$, which can be identified with the complex projective space
%corresponding to the Hilbert space $\mathcal{H}$. 

\subsection{Example}

Let us consider the manifold $\R^{2n}$ and the dynamical system
$$
\Gamma=p_j\frac{\partial}{\partial q_j}-\nu^2q_j\frac{\partial}{\partial p_j}.
$$

We can find alternative Hamiltonian structures connected to the factorization of 
$$
\Gamma=(q_j, p_j)
\left (
\begin{array}{cc}
 0 & -\nu^2 \\
1 & 0 
\end{array}
\right )
\left (
\begin{array}{c}
\frac{\partial}{\partial q_j} \\
 \frac{\partial}{\partial p_j}
\end{array}
\right )
=p_j\frac{\partial}{\partial q_j}-\nu^2q_j\frac{\partial}{\partial p_j}
$$
where
$$
\left (
\begin{array}{cc}
 0 & -\nu^2 \\
1 & 0 
\end{array}
\right )=
\left \{ 
\begin{array}{c}
\left (
 \begin{array}{cc}
 0 & -1 \\
1 & 0 
\end{array}
\right )
\left (
\begin{array}{cc}
 1 & 0 \\
0 & \nu^2 
\end{array}
\right )
\\
\left (
\begin{array}{cc}
 0 & -\nu^2 \\
\nu^2 & 0 
\end{array}
\right )
\left (
\begin{array}{cc}
 \frac 1{\nu^2} & 0 \\
0 & 1 
\end{array}
\right )
\end{array}
\right .
$$

In the case of $n=2$, by redefining position and momenta as $(q,p)\to (q',p)$,
according to the symmetric part of the decomposition we have: 
$$
\Gamma=
\nu \left ( p_j\frac{\partial}{\partial q'_j}-q'_j\frac{\partial}{\partial p_j} \right)
$$

Let us now consider in $\R^4$ the Hamiltonian function
$$
H=\frac 12 \nu \left ( p_1^2+q_1^2+p_2^2+q_2^2  \right),
$$
and the corresponding flow.

Therefore, $H=\frac 12 \nu$ implies $p_1^2+q_1^2+p_2^2+q_2^2=1$, i.e. a point in
the unit sphere.  A family of constants of the motion is provided by the set
$$
p_1q_2-q_1p_2 \quad p_1p_2+q_1q_2 \quad (p_1^2+q_1^2)-(p_2^2+q_2^2).
$$
It is simple to check that their Poisson bracket generate the Lie algebra of
$SU(2)$ and along with the dynamics they generate the Lie algebra of the full
unitary group $U(2)$. A foliation by two dimensional tori is given by the level
sets of 
$$
p_1^2+q_1^2, \qquad p_2^2+q_2^2,
$$
on the open dense submanifold where they are independent.

We notice that alternative Hamiltonian descriptions for $\Gamma$ may be related to
the immersion of $\Gamma$ into different Lie groups or Lie algebras. For instance,
we have seen that we may generate $U(2)$ when we use the Poisson bracket
defined as 
$$
\{q_a, p_b\}=\delta_{ab} \quad \{q_a,q_b\}=0=\{p_a, p_b\}.
$$
But if instead we consider
$$
\{q_1, p_1\}=1, \quad \{q_2,p_2 \}=-1, \quad \{q_a, q_b \}=0=\{p_a, p_b \};
$$
the same constants of the motion will generate the Lie algebra of the group
$SU(1,1)$. This would correspond to the case of the factorization of the matrix
$A$ associated with the linear dynamics into a skew-symmetric and a symmetric
one which is not degenerate but it is not positive definite.

It is also possible to consider the dynamics as part of an infinite dimensional
Lie algebra (deformations).  

In general we look for factorizations of the vector field by means of a Poisson
tensor and a differential one form ($\Gamma=\Lambda(dH)$), in this way both $\Lambda$ and $dH$
need not  be associated with matrices. In particular if we consider the
$(1,1)$--tensor field 
$$
T=(dp_j\otimes \frac{\partial}{\partial q_j}-dq_j\otimes \frac{\partial}{\partial p_j}),
$$
with any constant of the motion of the form
$$
F=F(p_1^2+q_1^2, p_2^2+q_2^2,p_1q_2-q_1p_2, p_1p_2+q_1q_2),
$$ 
we can associate an invariant one form \cite{MorFerLoVecMarRu:1990}
$$
\theta_F=d_TF.
$$
By a proper choice of the function $F$, we can make the two form $dd_TF$ non
degenerate, i.e. we can choose $F$ in such a way that the two form 
$$
\omega_F=dd_TF,
$$
is an invariant symplectic two form.

Moreover
$$
L_\Gamma d_TF=i_\Gamma dd_TF+di_\Gamma d_TF=0,
$$
implies that
$$
i_\Gamma d d_TF=-d(i_\Gamma d_TF),
$$
with 
$$
i_\Gamma d_T F=dF(T(\Gamma))=-\nu dF(\Delta)=-\nu L_\Delta F
$$
because $T(\Gamma)=-\nu\Delta$,  $\Delta$ being the dilation vector field.

 Any other invariant $(1,1)$--tensor field will give rise to alternative
 symplectic structure with the same procedure.

More generally, if $X_j$ and $X_{j+n}$ are pairwise commuting vector fields
$[X_j, X_k]=0$ $\forall j,k=1, \cdots, 2n$, with 
$$
L_{\Gamma}X_j=X_{j+n} \qquad L_\Gamma X_{j+n}=-X_j;
$$
the one form
$$
\theta_F=\frac 12\sum_j L_{X_j}d(L_{X_{j+n}}F)-L_{X_{j+n}}d(L_{X_j}F)
$$
is $\Gamma$--invariant and under proper choices of $F$ provides a symplectic structure
$$
\omega_F=d\theta_F.
$$
In addition, if we set
$$
Q_j=L_{X_j}F \qquad P_j=L_{X_{j+n}}F,
$$
we find a new linear structure and a new K\"ahler structure invariant under $\Gamma$:
$$
L_\Gamma Q_j=P_j \qquad L_\Gamma P_j=-Q_j.
$$
The new Hamiltonian function is given by
$$
\frac 12 \left ( (L_{X_j}F)^2+(L_{X_{j+n}}F)^2 \right)=\frac 12 \sum_j (Q_j^2+P_j^2),
$$
for any constant of the motion $F$, which makes $\omega_F$ not degenerate.

This construction of alternative K\"ahlerian structures applies therefore to the case of
quantum systems along the lines of \cite{ErIbMarMor:2007}.

By using the Lie algebra generated by 
$$
p_j, q_j, \mathbb{I},
$$
we can consider a larger algebra by using an infinitesimal generator of an
element in the automorphism group of the algebra itself
$$
\frac 12 \sum_j\left ( p_j^2+q_j^2\right).
$$
In general, any quadratic Hamiltonian may be considered as an element of the
automorphism group, the inhomogeneous symplectic group. In particular we may
find subalgebras or 
subgroups which contain just one quadratic Hamiltonian and the full linear
part. This would be a generalization of what is known as the oscillator group,
which is four dimensional \cite{streater:1967}. 

It is interesting to notice that the oscillator group allows for a
``contraction'' by letting $\nu\to 0$ and therefore it can reproduce the free
particle case. On the other hand, considering the unfolding procedure described
in
\cite{DavanMar:2005,DavanMarVal:2005}
it is also possible to describe the Hydrogen atom 
(Kepler problem) by using a nonlinear Harmonic oscillator (i.e. by introducing
an energy-dependent period to be able to accommodate for the energy-period theorem
when  dealing with the central-force problem). 
\section{The notion of independence}

\subsection{Proposal based on the geometric framework}
Let us go back again to the problem discussed above originated by the von
Neumann theorem, but from a more constructive perspective. Let us see how the
tools we introduced within the geometric description of quantum mechanics allow
us to introduce a notion of independence which may be convenient for our purposes.

Let us consider an even dimension vector space $V$ and a linear differential
equation associated to an operator $A$:
$$
\dot x=A x \qquad x\in V
$$

$A$ is defining a vector field on $V$. We can consider:
\begin{itemize}
\item there is a symplectic form $\omega$ on $V$ such that the vector field is
  Hamiltonian. 
\item Solve the inverse problem of Hamiltonian mechanics to find invariant symplectic
  forms
\end{itemize}

If $V$ is finite dimensional, we can obtain metric tensors (by using the
Hessian of the Hamiltonian function)  such that the
vector field is a Killing vector field for $g$. By proper rescaling the metric
tensor defined in this way, we may combine $\omega$ and the rescaled $g$ to define
an Hermitian product on $V$ which is preserved by the Hamiltonian
vector field. This allows us to introduce a set of quadratic functions on
$V$:
$$
B\to f_B=\langle x, Bx\rangle, 
$$
associated with Hermitian matrices with respect to the newly defined Hermitian
products. 

The set of (quadratic) functions can be endowed with different algebraic structures:
\begin{itemize}
\item the pointwise product
\item the Poisson bracket associated to the symplectic form
\item the non-local product which translates the product of operators at the
  level of quadratic functions:
$$
AB\to f_A\star f_B
$$
\item the skew-symmetric part of the above, which encodes the commutator
  algebra of the operators:
$$
f_A\star f_B-f_B\star f_A
$$
\end{itemize}

Let us now discuss how to introduce a concept of
functional independence. We have two possibilities:
\begin{itemize}
\item to consider the independence in the sense of the pointwise product and to
  use the standard differential calculus. Two functions are then said to be
  independent iff
$$
df_A\land df_B\neq 0 
$$ 
\item to consider independence with respect to the $\star$ product: in that case a
  non-commutative differential calculus is required and the issue becomes much
  more involved (see \cite{Segal2,Segal1,BiMarStern:2000,MarViZam:2006}).
\end{itemize}

In the finite dimensional case it is simple to see that the notion associated
with the pointwise product is convenient and provides a reasonable
framework. Besides it has the advantage of being easily extensible to the case
of infinite dimensional systems, once:
\begin{itemize}
\item a metric tensor compatible with the symplectic form has been defined. The
  definition can be taken from \cite{MarMorSimVen:2002}, where the operator $A$ is
  used: once the symplectic form has been chosen the Hamiltonian quadratic
  function $f_A$  is considered. This is a quadratic function, and hence it can
  be used to define a metric tensor

Then, there are two possibilities:
\begin{itemize}
\item it is positive definite: in this case it is used directly to write the
  metric tensor as
$$
g=
\left (
\begin{array}{cc}
dq &  dp
\end{array}
\right ) 
Hess f_A
\left (
\begin{array}{c}
dq \\
dp
\end{array}
\right )
$$
\item if it is not positive definite it can be proved (\cite{mascsmven:2005}) that the
  space $V$  can be decomposed into subspaces  (say $V_1$ and $V_2$) so that
  the functions may be   written as
$$
f_A=f_{A_1}-f_{A_2}
$$
where $A_1$ is a matrix with entries only on $V_1$ and $A_2$ with entries only
on $V_2$. A pseudo-metric (non definite metric) is thus obtained. The corresponding
  Hermitian structure will be pseudo-Hermitian on the total Hilbert space, but
  integrability may be then discussed on each Hilbert subspace where the metric
  may be considered to be Hermitian. 
\end{itemize}

\item the metric tensor allows to define suitable functions corresponding
  to the powers of the operator $A$. This is not a trivial issue, since
  depending on the domain  of $A$, the powers need not have an associated
  function (the corresponding quadratic expression need not be
  convergent). Hence the freedom  in the choice of the metric may be used to
  force all powers of the operator $A$ to be trace-class.
\end{itemize}

\subsection{An example}

  Let us consider an application of this construction to a particular
  example. Consider a finite level system defined on $\mathcal{H}$ and
  an operator $A$ and its square $A^2$. It seems obvious to consider
  both operators as dependent, from the point of view of
  the associative operator product.  But let us consider their independence
  from the point of view of the associated quadratic
  functions with respect to the differential calculus associated with their
  commutative product. Consider therefore  
$$
f_A=\langle\psi|A|\psi\rangle \qquad f_{A^2}=\langle\psi|A^2|\psi\rangle
$$

If the operator $A$ is self-adjoint with simple eigenvalues, we can consider
the basis of eigenstates, and write:
$$
f_A=\sum_j \lambda_{j}\bar z_jz_j \qquad f_{A^2}=\sum_j\lambda_j^2\bar
z_j z_j
$$

If we compute the differential of each function:
$$
df_A=\sum_j \lambda_j(z_jd\bar z_j+\bar z_jdz_j) \qquad df_{A^2}=\sum_j
\lambda_j^2(z_jd\bar z_j+\bar z_jdz_j) 
$$
And these two one forms satisfy
$$
df_{A}\land
df_{A^2}=\sum_{jk}(\lambda_j\lambda_k^2-\lambda_j^2\lambda_k) \bar z_jz_kdz_j\land d\bar z_k)=
\sum_{jk}\lambda_j\lambda_k(\lambda_k-\lambda_j) \bar z_jz_kdz_j\land d\bar z_k)
\neq 0
$$

Thus we see that, in the generic case, these two functions can be
considered to be functionally independent. The condition for this, in general,
will depend on the spectrum.

{\bf Remark}

We try to explain in a nutshell the main ideas involved in previous
constructions. 

Any algebra $A$ may be considered as a vector space $V$ along with a binary
product $B:V\times V\to V$. Given any linear vector space $V$, we may ``geometrize''
it by considering the embedding  $V\hookrightarrow \mathrm{Lin}(V^*, \R)\subset \F(V^*)$, i.e. any
element $u$ of $V$ is thought of as a linear function on the dual space $V^*$. If
$\alpha\in V^*$, we have 
$$
\hat u(\alpha)=\alpha(u), 
$$
i.e. $u$ is mapped into $\hat u$. Once we identify $\hat u$ with a function on
$V^*$, it makes sense to consider polynomial functions and construct more
general functions by means of the Weierstrass theorem.  On this space of
functions we may develop a standard exterior differential calculus. 

On the other hand, the binary bilinear product available on $V$, allows to
define a product on $\hat V$, by setting
$$
(\hat v\star \hat u)(\alpha)=\alpha(B(v,u)).
$$ 
This product is non-local and would require a non-commutative exterior
differential calculus.

To clarify our statements, let us consider the algebra of $n\times n$ matrices
$M(n,\R)$. We may consider its dual space generated by
$$
\{ e_{jk}=|e_j\rangle\langle e_k|\}, 
$$
constructed out of an orthonormal basis in some $n$--dimensional vector
space $E$. Clearly, 
$$
e_{jk}(A)=\langle e_k, Ae_j\rangle
$$
defines the dual pairing. We may define thus 
$$
\hat A(e_{jk})=e_{jk}(A),
$$
and the products

$$
(\hat A\cdot \hat B)(e_{jk})=e_{jk}(A)\cdot e_{jk}(B)
$$
and
$$
(\hat A\star \hat B)(e_{jk})=e_{jk}(A\cdot B).
$$

The first one is the usual Hadamard (or Shur) product among matrices and is
commutative and local. The exterior differential calculus associated to it is
the standard one on $\R^{n^{2}}$. 

The second product is the standard row-by-column product, and is non-local and
non-commutative. 

Thus, in a sense, the first product is originated only from the vector space
structure of the algebra, while the second one is specific of the binary
bilinear product which defines it.  Our suggestion is thus to use the exterior
differential calculus associated with the first (local and commutative) product
to define the notion of functional independence of observables in Quantum Mechanics.

%Free motion

\section{Examples: Harmonic oscillators and their deformations}

        From our previous discussion on classical integrability,  it is obvious
that Harmonic oscillators of any dimension will provide examples of integrable
systems, also in the quantum framework. We will use them as a reference model
to which all other quantum integrable systems should be related (those having a
discrete spectrum, of course).

Now we are going to review some properties of the system and we will present
then two examples of quantum integrable systems which exhibit the properties we
have discussed above: the deformed Harmonic oscillator and the Coulomb problem
in two and three dimensions.

\subsection{The standard Harmonic oscillator}
        As a preliminary simple example, we consider
the Harmonic oscillator. The Hamiltonian is well known:
$$
H=\frac 12 (P^2+Q^2)
$$
supposing $\hbar=1=\omega$.

        The dynamics is linear and can be integrated easily. This system is
obviously integrable from our point of view, and in  both schemes (vector
fields or operator subalgebras) the mapping $\Phi$ is the identity.

        To deal with the Harmonic oscillator one introduces
creation and annihilation operators $a, a^\dagger$ and the vacuum state $|0\rangle$ and
builds the states of the Fock space as

$$
|n\rangle=\frac{(a^\dagger )^n}{\sqrt{n!}}|0\rangle.
$$
These are eigenstates of the number operator $\hat n=a^\dagger a$, and thus define
the eigenstates of the Hamiltonian which is written as
$$
H=\hat n+\frac 12.
$$

The operators satisfy the commutation relation
$$
[a, a^\dagger]=\mathbb{I}
$$
and the number operator $(n=a^+a)$ thus generating
together with the identity, the well known Heisenberg-Weyl group ($H_4$).

The Hilbert space structure is fixed by the usual scalar product
$$
\langle n|m\rangle=\delta_{nm}.
$$

We notice that in the geometric framework these two objects determine the
Riemannian and Poisson structures of our description.

In the Heisenberg picture the equations of the motion are written as
$$
\dot a=-ia \qquad \dot a^\dagger =ia^\dagger 
$$

It is well known that the number operator classifies the irreducible
representations, as the Casimir corresponding to the subgroup $U(1)\subset
H_4$. The corresponding Hilbert space is the well known Fock space $V$.

Our aim is to show how the same dynamics can be described on the space of states
endowed with different Hermitian structures (therefore on 
different Hilbert spaces). The two different structures are related by a
nonlinear transformation which shows how  these types of
transformations arise within the quantum framework as described in
\cite{manko97:_f,manko97:_wigner_probl_and_alter_commut}.   

\subsection{A non trivial combination: the $f-$ deformed Harmonic
oscillator}

      Let us consider a construction similar to the previous case, but with
some novelties. Let us introduce the operators
$$
A=af(\hat n) \qquad A^\dagger =f(\hat n)a^\dagger ,
$$
where $f$ is a certain smooth function of the number operator (or equivalently,
a function of the energy of the system). As the number operator is a constant
of the motion for the dynamics of the Harmonic oscillator, it is clear that the
dynamics in this new operator basis has the same form:
$$
\dot A=-iA \qquad \dot A^\dagger =iA^\dagger 
$$ 

Associated with these operators we can also define a set of vector states (the
vacuum state $|0\rangle$ is the same as before):

$$
|N\rangle=\frac{(A^\dagger )^n}{\sqrt{n!}}|0\rangle,
$$
which are the eigenstates of a number operator defined as
$$
\hat N=A^\dagger A.
$$

A scalar product on this new set is defined also as
$$
\langle N|M\rangle=\delta_{NM}.
$$

We see thus that there is a one-to-one correspondence between the two sets of
states, since the mapping
$$
\Pi:|k\rangle \to |K\rangle \qquad k, K \in \N_0 
$$
is clearly bijective. It is important to remark, though, that the linear
structures are not preserved by this mapping, which is nonlinear,
i.e. $\Pi( |n_1\rangle+|n_2\rangle)\neq \Pi(  |n_1\rangle)+\Pi( |n_2\rangle)$. We shall come back to this point in the
next section. 

We have thus two different Hilbert space structures associated with the initial
system. We have here a situation very similar to the classical one: the
``abstract'' Heisenberg-Weyl algebra is realized in different manners and gives
rise to algebras of operators related by an intertwining operator which is not
a linear transformation. 

The situation can be summarized as follows:

\begin{itemize}
\item With respect to the initial Hilbert space structure, we can write the
  commutation relation of $a$ and $a^\dagger $ as
$$
\langle n, [a, a^\dagger]m\rangle=\delta_{nm}. 
$$
An analogous relation exists for the operators $A$ and $A^\dagger$ with respect to
the new Hermitian structure:
$$
\langle N, [A, A^\dagger]M\rangle=\delta_{NM}. 
$$
\item But if we consider $A$ and $A^\dagger$ as operators acting on the original Fock
  space, the commutation relations with respect to this action
  have the following form
$$
\langle n, [A, A^\dagger]m\rangle=((n+1)f^2(n+1)-n f(n)^2)\delta_{nm}. 
$$
We can write thus that, with respect to the original Fock space structure, 
$$
[A, A^\dagger]=(n+1)f^2(n+1)-n f(n)^2\mathbb{I}
$$
\end{itemize}

However the equations of motion of the Harmonic oscillator, i.e. the vector
field,  are the
same in both  cases, but the choice of one Hilbert space structure or the other
makes the description quite different when written in the ``wrong''
framework. And notice also that 
both Hilbert space structures are related to each other  by a
transformation which is nonlinear. The corresponding transformations at the
level of the symplectic structures (i.e. the transformations which relate the
commutation relations) define a non-canonical nonlinear transformation.

We can give some more details of the construction if we choose a particular
example of the mapping $f$:
\subsubsection{One dimensional example}
A typical example of this construction is provided by the so-called
$q$--deformed Harmonic oscillators, defined by the expressions:

$$
n_q=\frac{\sinh n\hbar}{\sinh \hbar} \ \ \ q=e^\hbar  
$$
where $n$ is the undeformed number operator introduced above.  Analogously 
the creation-annihilation operators are written in terms of the undeformed
ones:
$$
a_q=af(n) \ \ \  a^+_q=f(n)a^+
$$
where 
$$
f(n)=\sqrt{\frac{n_q}n}
$$

Notice that this transformation is invertible, since we can write
$$
a=A\sqrt{\left ( \frac{\log (\hat N\sinh \hbar +\sqrt{\hat N^2\sinh^2\hbar+1}}{\hbar \hat
    N}\right )}. 
$$

But it is simple to read from here the non-linearity of the mapping $\Pi:|k\rangle\to |K\rangle$
introduced above, since the relation for the number operator is thus written
as:
$$
\hat n=\frac 1\hbar \log \left ( \hat N\sinh \hbar+\sqrt{\hat N^2\sinh^2\hbar +1 } \right ).
$$
The mapping $\Pi$ is just read from here since it is defined on the set of
eigenstates of these two operators ($\hat n$ and $\hat N$).

The expression for the new commutation relation in terms of the old Hilbert
space structure corresponds to
$$
[A, A^\dagger]=\left (\hat N(\cosh \hbar -1)+\sqrt{\hat N^2\sinh^2\hbar +1}\right )\mathbb{I}
$$

        For this scheme, we define the Hamiltonian of the $q$-deformed 
oscillator to be:

$$
H_q=\frac 1\hbar \log \left ( \hat N\sinh \hbar+\sqrt{\hat N^2\sinh^2\hbar +1 }+\frac 12\right ).
$$
With this operator and the original Hilbert space structure, we can write the
associated vector field:
$$
[A, H_q]=A,
$$
which gives the original dynamics written in terms of the new operators.

\subsubsection{Two dimensional case}

For the cases with more than one oscillator 
we have the option of keeping or not the $U(N)$ symmetry of the
undeformed problem. If we prefer to keep the symmetry, we define the 
deformed creation and annihilation operators as:
$$
a_q=af(n) \ \ \  a^+_q=f(n)a^+
$$
$$
b_q=af(n) \ \ \  b^+_q=f(n)a^+
$$
where now $n=n_a+n_b$
(to break it, one should choose for instance only one of the number operators for each
deformation).

        The Hamiltonian is now obtained as:
$$
H=\frac 12 [ nf^2(n)+(n+2)f^2(n+1)]
$$

which exhibits the kind of combination of models we have exposed above. It 
is a combination of Harmonic oscillators with a function which is a
``constant of the motion''  on each description. 

        If we want to break the symmetry and define a different combination
for the Hamiltonian, we define the creation and annihilation operators as:

$$
a_q=af(n_a) \ \ \  a^+_q=f(n_a)a^+
$$
$$
b_q=af(n_b) \ \ \  b^+_q=f(n_b)a^+
$$
in such a way that the Hamiltonian becomes:
$$
H=\frac 12 [n_af^2(n_a)+n_bf^2(n_b)+(n_a+1)f^2(n_a+1)+(n_b+1)f^2(n_b+1)
$$

        In this case is even more clear the way models are combined.
We have two different oscillators, and the construction of a Hamiltonian
with different factor for each representation of them.

        The mapping $\Phi$ relating the two descriptions, acts only on
        each particular representation of 
the $H_4$ algebra of each particular oscillator. It transforms the algebra and 
leaves invariant the Hilbert space, which is constructed as the product of 
both Fock spaces in the usual way. The global transformation is constructed
as:
$$
\Phi=\sum f(n_i)\Phi^i
$$ 
where $\Phi^i$ is the one dimensional transformation and it is equal to the
identity.

\subsection{The Coulomb problem}
As an example of the potential applications of the composition of models to the
study of quantum problems let us consider a simple example in the Schr\"odinger
picture.

        It is well known (see for instance \cite{coulomb}) that the system
may be described with the help of a combination of Harmonic oscillators for the
bound states in the
$D=2$ and $D=3$ cases. From our point of view this implies that the system
is integrable at the quantum level (for these dimensions and regimes).

        For both cases we decompose the Hilbert space in the eigenspaces
of the Hamiltonian, let us call them $U_{E_\alpha}$. Now, on each
subspace we perform a change of coordinates that maps $U_{E_\alpha}$
on a suitable subspace of a Fock space.:
\begin{itemize}
\item For the two dimensional case the change of coordinates is simply
the transformation into parabolic cylinder coordinates:
$$
 x=\frac 12 (u^2-v^2) \ \ \ \  y=uv 
$$
where the point $(-u,-v)$ and $(u,v)$ are identified.
This change transform the eigenvalue equation 
$$
-\frac 12 \left ( \frac{\partial^2\psi}{\partial x^2}+
\frac{\partial^2\psi}{\partial y^2}\right ) -
\frac k{\sqrt{x^2+y^2}}\psi =E\psi 
$$
into
$$
-\frac 12 \left( \frac{\partial^2\psi}{\partial u^2}+
\frac{\partial^2\psi}{\partial v^2}\right ) -
\frac 12\omega^2(u^2+v^2)\psi =E'\psi 
$$
where the wave-functions should verify $\psi(-u,-v)=\psi(u,v)$,
where $\omega^2=2|E|=\frac {2k}{n^2}$ and $E'=k=n\omega$
where $n$ has to be odd to satisfy the parity condition of the wave
functions. We construct thus a mapping from the Hilbert space of bound
states of the two dimensional Coulomb problem into the subspace of the
product Fock space $V_u\otimes V_v$ corresponding to an odd number of
particles. There is a dependence of the mapping on the Hamiltonian, which
produces a different frequency for each oscillator.
\begin{equation}
\Phi_{E_\alpha}:U_{E_\alpha}\to V_{uv}^{odd}\subset V_u\otimes V_v
\end{equation}
The complete mapping is the union of these $\Phi_E$ which are formally
identical:
$$
\Phi=\cup_\alpha \Phi_{E_\alpha}
$$

\item For the three dimensional case, the situation is analogous
(see \cite{coulomb} as the original reference) . Now
we make use of the $\goth{su}(2)\oplus \goth{su}(2)$ symmetry algebra of
the Coulomb problem (corresponding to suitable combinations of the
angular momentum and the Runge-Lenz vector if we define it in terms of
generators). We construct then a two dimensional Harmonic oscillator
associated with each of the $\goth{su}(2)$ group, thus obtaining a four
dimensional isotropic Harmonic oscillator.

         The change of coordinates is constructed in two steps:
\begin{eqnarray*}
x_1=\mu \nu \cos \phi \ \ \ x_2=\mu \nu \sin \phi \ \ \  x_3=\frac 12
(\mu^2-\nu^2) \\
\xi_\mu=\mu \cos \phi \ \ \ \eta_\mu=\mu \sin \phi \\
\xi_\nu=\nu \cos \phi \ \ \ \eta_\nu=\nu \sin \phi
\end{eqnarray*}
where again $\omega^2=2|E|$ (the change is different for each eigenspace
of the Hamiltonian). These changes 
transform now the system  (this time considered
in three dimensions) into:
$$H\psi=(H_\mu+H_\nu)\psi=E'\psi$$
where
$$H_\alpha=-\frac 12 \left( \frac{\partial^2}{\partial \xi_\alpha^2}+
\frac{\partial^2}{\partial \eta_\alpha^2}\right ) -
\frac 12\omega^2(\xi_\alpha^2+\eta_\alpha^2) $$
and $E'=k=\omega(n_\mu+n_\nu+1)$.

        The mapping should be understood now applying each eigenspace of the
Hamiltonian of the hydrogen atom, representation by representation of the
angular momentum and the Runge-Lenz vector, onto the corresponding 
representations of the Hamiltonian of the Harmonic oscillator. 

        We can write analogously to the previous case:
\begin{equation}
\Phi_{E_\alpha}:U_{E_\alpha}\to V_{\mu \nu}^{odd}\subset V_\mu\otimes
V_\nu
\end{equation}
where now $V_\alpha$ is the Fock space corresponding to a two dimensional
oscillator. For the whole Hilbert space we can write again:
$$\Phi=\cup_\alpha \Phi_{E_\alpha}$$  
\end{itemize}

A different approach, where starting from the classical situation the
construction of the quantum one is directly related to the geometry of the
Lagrangian description is provided in \cite{DavanMarVal:2005}.

\section{Conclusions and outlook}

In this paper we have pointed out various problems that a definition of quantum
integrability has to face. The main one seems to be the definition of
functional independence which, in our present understanding, requires a
differential calculus on commutative algebras.  This calculus deals with
derivations and is fully captured by the notion of differential manifold, be it
finite or infinite dimensional. 
For this reason we believe that the geometrization of quantum mechanics can be
a good starting point. In this approach we find that alternative products on
functions coexist: 
\begin{itemize} 
 \item a non-local product (on quadratic functions) which captures the essence
   of quantum mechanics in terms of indetermination relations and a rule for
   composing systems, 
\item a local commutative product which allows for a differential calculus in
  terms of derivations which are local. 
\end{itemize}

At the moment, it seems that the latter is useful, if not unavoidable, to state
functional independence of ``observations'' (quadratic function version of the
observables).  

Usual treatments of integrability (\cite{CasiDeVitoLev:1997,TemWin:2001})
at the quantum level deal with Hilbert spaces 
considered as spaces of square integrable functions on some ``configuration
space''. It seems that the notions we are advocating are provided by Lie groups
and their unitary representations, along with the associated representations of
their Lie algebras (and their enveloping algebras) on the set of analytic or
smooth vectors \cite{nelson,Davies:1971}.

The convolution product and the pointwise product available on the group seem
to be related to the two products we are identifying. All these considerations
appear very neatly in the so called Quantum Mechanics on phase space (the
Wigner-Weyl formalism, see \cite{CaC-GMar:2007b}) There the phase space is an
Abelian vector group and the 
convolution product (Moyal product) is associated  with its central extension
defined with the help of a symplectic structure). The algebra of differential
operators is identified as the algebra of operators acting on square integrable
functions defined on Lagrangian subspaces of this symplectic vector space.

The representation of the enveloping algebra allows us to deal with suitable
deformations (quantum groups and Yang Baxter relations). 

The transformation from one Lagrangian subspace to another one requires the
introduction of pseudo-differential operators. Nonlinearly related
actions of $\R^{2n}$ on the phase space, when considered as Weyl systems
\cite{ErIbMarMor:2007}, give rise to nonlinear transformations in the quantum
framework. 

Some of these statements need further elaboration to be able to convert them
into propositions and theorems. We hope to come back to these questions in the
near future.

Another spin-off of this geometrization procedure is the possibility of using
methods and conceptual constructions elaborated in the framework of classical
physics (vector fields, Poisson brackets, Riemannian vector fields) also in the
framework of quantum physics, with the additional intervention of  additional
structures which may be traced back to the existence of the fundamental
constant of nature provided by Planck's constant \cite{MarScoSimVen:2005}.

The difficulties we have to face can be traced back to the lack of a mature
mathematical treatment of infinite-dimensional differential manifold even
though very good books are already available
(\cite{Lang:2002,Cartan:1970,KoMichSlo:1993}) . Our
hope is to be able to apply in the very near future these methods to study
quantum control problems using as much as possible the experience available in
the ``classical framework''.

{\bf Acknowledgements}
This paper is an extended version of the seminar given by one of us (J. C-G) in
the framework of the Conference ``Mathematical Structure on Quantum Mechanics
II'' which took place in Madrid (Spain), in March 2008. We would like to thank
the organizers of that event for giving us the opportunity of presenting the
work and to the participants for their interesting comments and suggestions.

%\bibliographystyle{plain}
%\bibliography{quantumt}

\end{document}